\documentclass[10pt]{article}
\usepackage{amssymb}
\usepackage{amsfonts}
\usepackage{amsmath}
\usepackage{geometry}
\usepackage{pdflscape}
\usepackage{amsmath, amssymb, latexsym,amsthm}
\usepackage{graphicx}
\graphicspath{{./graphs/}}
\usepackage{comment}
\usepackage[colorlinks,linkcolor=blue,anchorcolor=blue,citecolor=blue]{hyperref}
\usepackage{setspace}
\usepackage{hyperref}
\usepackage{color}
\usepackage{bm}
\usepackage{bbm}
\usepackage{palatino}
\usepackage{caption}
\usepackage{booktabs}
\usepackage[flushleft]{threeparttable}
\usepackage[numbers,sort&compress]{natbib}
\setcitestyle{round}

\usepackage{tabularx}  
    \usepackage{dcolumn} 
    \usepackage{csvsimple}  
    \newcolumntype{d}[1]{D{.}{.}{#1}}
    \usepackage[flushleft]{threeparttable} 
    \usepackage[labelfont=bf, font=footnotesize, labelsep=period]{caption}
    \usepackage{subcaption}
    \usepackage{graphicx}   
    \usepackage{scalerel,stackengine}
    \stackMath
    \newcommand\reallywidehat[1]{%
    \savestack{\tmpbox}{\stretchto{%
    \scaleto{%
        \scalerel*[\widthof{\ensuremath{#1}}]{\kern.1pt\mathchar"0362\kern.1pt}%
        {\rule{0ex}{\textheight}}
    }{\textheight}%
    }{2.4ex}}%
    \stackon[-6.9pt]{#1}{\tmpbox}%
    }
\usepackage{float}
\usepackage{pgfplots}
\usepackage{pgfplotstable}
\usepackage{tikz}
\usetikzlibrary{decorations.pathreplacing,angles,quotes}
\usetikzlibrary{positioning, matrix}
\usetikzlibrary{shapes,arrows}
\usepackage{adjustbox}
\usepackage{ulem}
\usetikzlibrary{calc}
\usepackage{multicol}
\usepackage{multirow,array}
\usepackage[utf8]{inputenc}

\newcolumntype{K}{>{\centering\arraybackslash}m{4cm}}
\newcolumntype{L}{>{\centering\arraybackslash}m{4cm}}

\usepackage{footmisc}      

\title{Getting Explicit Instruction Right}

\author{%
  Richard Holden\footnote{Holden: UNSW Business School, \href{mailto:richard.holden@unsw.edu.au}{richard.holden@unsw.edu.au}.}%
  \and
  Fabio I. Martinenghi\footnote{Martinenghi: Corresponding author, University of Newcastle, 
  \href{mailto:fabio.martinenghi@newcastle.edu.au}{fabio.martinenghi@newcastle.edu.au}. \\
  We are grateful to the New South Wales Department of Education, Gregor Pfeifer, and seminar participants at UNSW. 
  \textbf{Ethics approval}: University of New South Wales Human Research Ethics Advisory Panel (HREAP), Approval No.\ HC220103.
  }%
}
\date{\today}

\pgfplotsset{compat=1.18}

\begin{document}
	
	\maketitle
	
	\begin{abstract}
		  There has been substantial public debate about the potentially deleterious effects of the long-run move to ``inquiry-based learning'' in which students are placed at the center of an educational journey and arrive at their own understanding of what is being taught. There have been numerous calls for a return to ``direct'' or ``explicit'' instruction. This paper focuses on the impact of implementing explicit instruction via peer modelling, estimating its causal effect on student performance in standardized tests. We utilise a unique setting in Australia\textemdash a country in which all students in grades 3, 5, 7, and 9 undergo annual basic skills tests (``NAPLAN'')\textemdash and administrative data. We take a synthetic control and use Charlestown South Public School\textemdash a median-performing school\textemdash as a case study. Importantly, Charlestown South implemented explicit instruction via peer modelling, with teaching staff sitting-in during the classes of a high-performing explicit-instruction school. We find that the student performance gains are large (0.9-2.7 standard deviations) and persistent. 
		
		\medskip
		
		\medskip
		
	\end{abstract}
	
	\newpage

\section*{Significance Statement} 
Schools across advanced economies face declining performance and debate over effective pedagogy. We study a practical, scalable implementation of Explicit Instruction: teachers from a median-performing public school in New South Wales learned the approach by observing a top-performing school and then applied it school-wide. Using administrative test data and a synthetic control design, we estimate large and persistent gains in reading and numeracy for Years 3 and 5. These results provide causal evidence that a high-fidelity, peer-modelling implementation of Explicit Instruction can substantially raise achievement at scale in a public-school setting, informing policy makers seeking cost-effective ways to improve student performance.

\section{Introduction}
Over the last two decades, in many advanced economies, there has been a systematic decline in the academic performance of high school students. This is true, for instance, in the ``Programme for International Student Assessment'' (PISA) tests run by the Organisation for Economic Co-operation and Development (OECD), which tests the skills of 15-year-olds in  mathematics, science and reading. In these tests, countries like the United States, United Kingdom, and Australia have seen sharp declines in absolute and relative terms.

While the source of this decline is hard to pin down, many have speculated that a long-run trend in the way students are taught is to blame. This trend has involved a move away from teachers directly imparting learning to students, toward so-called ``inquiry-based learning'' in which students are placed at the center of an ``educational journey'' where they arrive at their own understanding of what is being taught. This approach emerged from the ``constructivist'' movement in education which emphasizes hands-on learning aimed at solving ``real-world'' problems. This has led to numerous calls for a return to ``direct'' or ``explicit'' instruction \citep{convo2019}. This is sometimes called a ``back-to-basics'' approach to education.

The purpose of this paper is not to litigate the cause of declining educational performance in many advanced economies. Rather, we focus on estimating the causal effect of implementing the explicit instruction pedagogy via peer-modelling on student performance in standardized tests. Here, by peer-modelling implementation, we mean the process where teachers from a given school (i) learn a pedagogical approach by sitting in the classes of teachers from a school successfully implementing such approach, and (ii) use the approach in their own school. To measure this impact, we utilize a unique setting in Australia\textemdash a country in which all students in grades 3, 5, 7, and 9 undergo annual basic skills tests (``NAPLAN'')\textemdash and apply the synthetic control method \cite{abadie03} to administrative data. We find strong causal evidence that a peer-modelling implementation of explicit instruction improves student performance and the magnitude is substantial, with point estimates ranging between 0.9 and 2.7 standard deviations. We emphasise the implementation aspect and build the counterfactual outcomes using a pool of schools that includes schools already attempting to introduce explicit instruction in their pedagogical approach. 

There is a large and multi-disciplinary literature providing evidence of positive sign on the effectiveness of explicit instruction and similar pedagogical approaches. 
While Explicit Instruction is defined in various ways throughout the literature, we follow the definition from \citeauthor{archer2010} \citep{archer2010}\textemdash a classic textbook on this topic. See Section \ref{sec:pillar} for details. 
The relevant Education literature include studies on reading for primary school students \citep{andreassen11}, on mathematics in Kindergarten \citep{doabler15}, and on using explicit instruction to teach mathematics to primary \citep{kroes2003} and secondary school students with special needs \citep{powell21}. For reviews focusing on the education literature, see \citeauthor{archer2010} \citep{archer2010}, \citeauthor{hughes2017} \citep{hughes2017}, and \citeauthor{Stockard2018} \citep{Stockard2018}'s comprehensive meta-analysis.

Economists have also been interested in assessing the impact of explicit and structured teaching approaches on student outcomes. Difference‐in‐differences studies of the ``Literacy Hour'' and synthetic-phonics reforms find substantial gains in early reading outcomes (approx. $0.2-0.3$ standard deviations; s.d.) and long-run benefits for low-performing students \citep{MachinMcNally2008,Machin2018}. Using field experiments in the US, \cite{Fryer2014} provides evidence that importing the direct instruction practices of charter schools\textemdash extended instructional time, data‐driven teaching, intensive tutoring\textemdash into public schools raises mathematics achievement by up to 0.18 s.d. one year after the implementation. \citeauthor{SchwerdtWuppermann2011} \citep{SchwerdtWuppermann2011} estimates a fixed-effects model on US data and finds that approaches where the teacher takes centre stage and presents the content in a structured format outperform problem‐solving approaches. \citeauthor{JacobLefgren2004} \citep{JacobLefgren2004} uses a regression discontinuity design and focusing on mathematics, and \citeauthor{Banerjee2007} \citep{Banerjee2007} and \citeauthor{Piper2018} \citep{Piper2018} running RCTs in India and Kenya, respectively, to study the effect of structured approaches on numeracy and literacy skills\textemdash all finding positive impacts. 
Other examples of studies related to structured-teaching interventions are \citeauthor{RomeroSandefur2022} \citep{RomeroSandefur2022}, an RCT where 93 Liberian schools outsourced were randomly-assigned to be managed by private operators\textemdash who used scripted, highly-structured curricula\textemdash and \citeauthor{DufloKiesselLucas2024} \citep{DufloKiesselLucas2024}, where where one of the treatment arms of their RCT in Ghana was based on remedial tutoring that uses structured lesson guides. They also find positive impacts.

Our paper adds to this literature in several ways. 
First, it shows the importance of a high-fidelity implementation of \citeauthor{archer2010} \cite{archer2010}'s Explicit Instruction approach. 
In our case study, this is achieved by the teachers of the treated school emulating the teachers of a successful Explicit-Instruction school.
Specifically, the principal and the teachers of the treated school visited the successful school for three days and learned how to best implement Explicit Instruction by sitting-in during their classes (called ``modelling'', in the education literature).
This seems not only to be an effective solution, but also a cost-effective one, compared\textemdash for instance\textemdash to hiring dedicated coaching staff for long periods of time (see Section \ref{sec:other_attempts}), and particularly in high-income countries like Australia, where the cost of labour is high. 
The fact that this is happening outside of an experimental setting\textemdash and within a public-school context\textemdash should also reassure policy-makers that the performance gains we estimate are attainable. 

Second, our intervention of interest does not target a specific subject \citep[e.g., ][]{MachinMcNally2008, doabler15, Machin2018} or student sub-group \citep[e.g., ][]{torgesen2010, powell21, CortesGoodmanNomi2015}, instead changing the school pedagogy across all teaching areas and benefiting all students. This allows our evidence to provide insight on the broad benefits of a school-wide adoption of Explicit Instruction, compared to a partial or targeted one. Studies of charter schools and their practices \citep[e.g., ][]{DobbieFryer2013, Fryer2014} are similar to our study in this respect, but do not focus strictly on features of Explicit Instruction. 

Third, we are able to follow students for 8 years after the implementation, while even most studies with longer follow-up times are limited to less than two years \citep[e.g., ][]{JacobLefgren2004, Banerjee2007, torgesen2010}. This allows us to (i) learn that transitioning from one pedagogical approach to another can take time, and (ii) that that the benefits of Explicit Instruction are persistent in the long run. \citeauthor{MachinMcNally2008} \cite{MachinMcNally2008} follow students between 1996 and 1998, but focus exclusively on impacts of reading and English skills. 

Finally, there is a paucity of evidence on how well this approach performs with respect to large-scale standardized tests. While \citeauthor{oliver2021} \cite{oliver2021} and \citeauthor{jerrim2023} \cite{jerrim2023} use PISA Science scores as outcomes, the causality of their findings purely relies on observed confounders (selection on observables).

\subsection*{Synthetic control method}
While the Synthetic Control Method (SCM) was first introduced by \citeauthor{abadie03} \cite{abadie03} to study
the effect of conflict on economic output, since then it has become a mainstream method in any field working on causal questions using observational data \citep[e.g.][]{Bruhn2017, Roopsind2019, Bayer2020, Mitze2020, West2020, Yan2021, GarciaBulle2022, Scheuer2022, Wang2022, Milkman2022, West2023, Wu2023}.

We compare Charlestown South to a ``synthetic twin'' built as a weighted average of similar NSW schools that did not adopt the peer-modelled Explicit Instruction. The weights are chosen so that, before 2014, the synthetic twin closely matches Charlestown South on test-score trajectories and key school characteristics. After 2014, any systematic divergence in outcomes between Charlestown South and its synthetic twin is interpreted as the causal effect of the implementation. We also conduct standard placebo tests by pretending that other schools were treated to assess whether effects of similar magnitude would appear by chance.

We provide further methodological information in Section \ref{sec:method} and in the appendix. 

\section{Background}
\subsection{The Pillars of Explicit Instruction\label{sec:pillar}}
While \cite{archer2010} define sixteen Explicit Instruction principles,
we report the five essential components identified in \citeauthor{hughes2017} \cite{hughes2017}.
The first essential component consists in \textit{segmenting complex
    skills}, where complex tasks are divided into smaller simpler units.
The second is to indicate important features of the content by either
\textit{showing}, if an action is being taught, or \textit{telling}, i.e. thinking aloud, if a concept is being taught.

The third is promoting successful engagement with the help of
\textit{prompts that are gradually withdrawn}. An example of this is providing suggestions or partial solutions to students solving an exercise which decreases with each exercise iteration.
The fourth component is \textit{frequently querying and engaging} with students, which gives the teacher a  chance to provide immediate feedback and to monitor how well students are understanding the content.
Finally, the fifth component is to create \textit{practice} opportunities,
especially if paired with affirmative or corrective feedback.

Note that, while some components of Explicit Instruction
are found in other pedagogical approaches, Explicit Instruction is
distinct from similar-sounding pedagogies such as ``Direct Instruction'' and
``Direct Explicit Instruction'' \citep{hughes2017}. For instance,
Direct Instruction includes scripted lessons\textemdash which are
absent in Explicit Instruction. Moreover, Direct Explicit Instruction and
Direct Instruction include both curricular (what to teach) and instructional
(how to teach it) elements, while Explicit Instruction purely focuses on instruction
\citep{hughes2017}.

\subsection{The Charlestown South Case Study}
In an interview conducted as part of another research project, the principal of Charlestown South, a primary (K-6) school in the Newcastle area, New South Wales (NSW), explained how in 2014 he and many teachers had become dissatisfied with the academic performance of their Year-6 students, who scored much lower than their peers from other neighboring schools in the entry test to a local high school.

He sought out and found the school with the highest NAPLAN scores in Australia\textemdash a private school in Melbourne, Victoria. After making contact, he and several of his teachers visited the school. During the visit, they learned about the successful use of Explicit Instruction. The teachers had the chance to see this approach in action and once back at Charlestown South they began implementing it\textemdash although he reports that it took them around two years of work to fully grasp it and implement it correctly.

The principal emphasized the importance of a particular feature
in their version of Explicit Instruction. Namely, the review and practice of the content of the previous lesson before the start of a new lesson, in what are often called ``warm-ups''. This practice is encompassed by the 6th of the 16 elements of Explicit Instruction presented in \citeauthor{archer2010} \cite{archer2010}. It serves the purpose of moving learned content from the short-term to the long-term memory. It also helps highlight the cumulative dimension of learning\textemdash how each new lesson builds upon the previous one.

\subsection{Other attempts at introducing Explicit Instruction in NSW \label{sec:other_attempts}}

Before Charlestown South started implementing Explicit Instruction, 
the New South Wales public-school system had already 
introduced several structured-teaching programmes that tried to introduce some principles of Explicit Instruction.  
We provide some examples below.

The \textit{Early Action for Success} (\textsc{EaFS}) policy was launched 
in 2012 across the Kindergarten, Year 1 and Year 2 classrooms 
of 228 low-performing schools. 
It provides target schools with ``instructional leaders'', 
specially-hired teachers whose role is to train teacher staff 
in implementing explicit instruction routines both targeting the whole 
classroom and small groups of students who are not 
performing well 
 \citep{cese_eafs_2016}.  
The \textit{Language, Learning and Literacy} programme (\textsc{L3}) was piloted in
2010 and scaled to more than 900 Kindergarten classrooms by 2014, providing
teachers with scripted, thirty-minute sequences on guided reading and writing 
\citep{nsw_l3_2015}.  
The \textit{Focus on Reading} was rolled 
out state-wide in 2011 and trained teachers Years 3–6 teachers turn 
comprehension lessons into a deliberate, step-by-step routine. 
\citep{nsw_for_2013}.  All these initiatives are in line with 
research on evidence-based best practices in teaching conducted by 
the NSW Department of Education's Centre for Education Statistics and 
Evaluation \citep[CESE; ][]{nsw2013great}, which 
recommended explicit/direct teaching approaches based on 
a review of the literature and was the basis of the 
change in pedagogical approach in NSW public schools marked 
by \citeauthor{{gtinl_2013}} \cite{{gtinl_2013}}.  

All the above is relevant for our study as it implies that the counterfactual 
in our evaluation is not an inquiry-dominated system but one in which there are 
competing\textemdash and unevenly distributed\textemdash attempts at implementing 
the Explicit Instruction approach.

\section{Data}
We use school-level data between 2010-2021 provided by the NSW Department of Education and covering all NSW public schools. We focus on primary (K-6) schools, since Charleston South is a primary school. Our four outcomes of interest are Year-3 and Year-5 NAPLAN scores in Numeracy and Reading. The NAPLAN test was not conducted in 2020 due to the coronavirus pandemic. Thus, all data associated with year 2020 was dropped. To the NSW Department of Education data, we link yearly postcode-level mean individual-taxable-income data, as recorded by the Australian Taxation Office.

The NSW Department of Education dataset gives us information on: each school's geo-location (both encoded as latitutde-longitude coordinates and as postcodes), whether it is academically selective, how remote is its surrounding area, whether it includes at least one Opportunity Class (a form of ``tracking''), the year in which it hired the first teacher, school-level data on student attendance, the school's type (primary or secondary) and its sub-type (e.g. Kindergarten to Year-6), whether it is single-sex or coeducational, if it is open until late in the  and if it has a healthy canteen. It also reports time-varying school-level
information such as the Index of Community Socio-Educational Advantage (ICSEA)
\footnote{
    The ICSEA is an index that summarizes
    information about the occupation and earnings of children's parents within a given school, how remote the school is and the percentage of indigenous student enrolled.
}, the attendance rate, the share of girls, the share of indigenous students, the share of students with a language background other than English, the average teacher-student ratio.

\section{Results}
For each of our four outcomes, we construct a distinct synthetic control, with its associated
set of weights.\footnote{We produce our estimates and plots using the scpi R package \citep{cattaneo2022scpi}.
} However, since the four sets of weights estimated are identical to each other after rounding them to two decimals, we treat them as one in our discussion. As shown in Table \ref{tab:weights}, Synthetic Charlestown South's biggest ``donor schools'' are the Charlestown
(38\%) and Hillsborough (36\%) public schools\textemdash both in the vicinity of Charlestown South. This is because one of the features we matched on was ``radial proximity to Charlestown South''.

\begin{table}[h]
    \centering
    \begin{threeparttable}
        \caption{Schools used to construct Synthetic Charlestown South}
        
\begin{tabular}[t]{lr}
\toprule
School name & Weight\\
\midrule
Charlestown Public School & 0.381\\
Hillsborough Public School & 0.361\\
Carrington Public School & 0.098\\
Biraban Public School & 0.058\\
Stockton Public School & 0.047\\
Farmborough Road Public School & 0.034\\
Merewether Public School & 0.016\\
Coledale Public School & 0.006\\
\bottomrule
\end{tabular}

        \begin{tablenotes}
            \footnotesize
            \item \textbf{Note.} This table presents the weights used to
            construct Synthetic Charlestown South.
            They sum up to one, although they do not appear to
            due to rounding. Once rounded, the weights used are identical
            across our four outcomes. This is a special case and it
            is not true in general.
        \end{tablenotes}
        \label{tab:weights}
    \end{threeparttable}
\end{table}

In Table \ref{tab:balance}, we compare some key features of Charlestown
South with those of its synthetic counterpart. The two schools are similar in
all features but for ``year of first teacher'' (50 years of difference),
a date that marks the year in which the school became operational.

\begin{table}[h]
    \centering
    \begin{threeparttable}
        \caption{Balance between Charlestown South and Synthetic Charlestown South}
        
\begin{tabular}[t]{lrr}
\toprule
  & Charlestown South & Synthetic Charlestown South\\
\midrule
Attendance (share) & 0.95 & 0.93\\
Enrolments (FTE) & 191.36 & 172.51\\
ICSEA & 1099.73 & 1074.04\\
Postcode mean taxable income & 66809.24 & 71940.85\\
Females (share) & 0.46 & 0.48\\
\addlinespace
Mean class size & 24.02 & 23.23\\
Non-English language background (share) & 12.00 & 11.64\\
Radial distance for Charlestown South & 0.00 & 48.31\\
Year of first teacher & 1963.00 & 1913.29\\
\bottomrule
\end{tabular}

        \begin{tablenotes}
            \footnotesize
            \item \textbf{Note.} This table presents the mean value for each
            covariate used in the synthetic control analysis.
        \end{tablenotes}
        \label{tab:balance}
    \end{threeparttable}
\end{table}

\begin{figure}
    \begin{subfigure}{.5\textwidth}
        \centering
        \caption{Numeracy Year-3}
        \includegraphics[width=\textwidth]{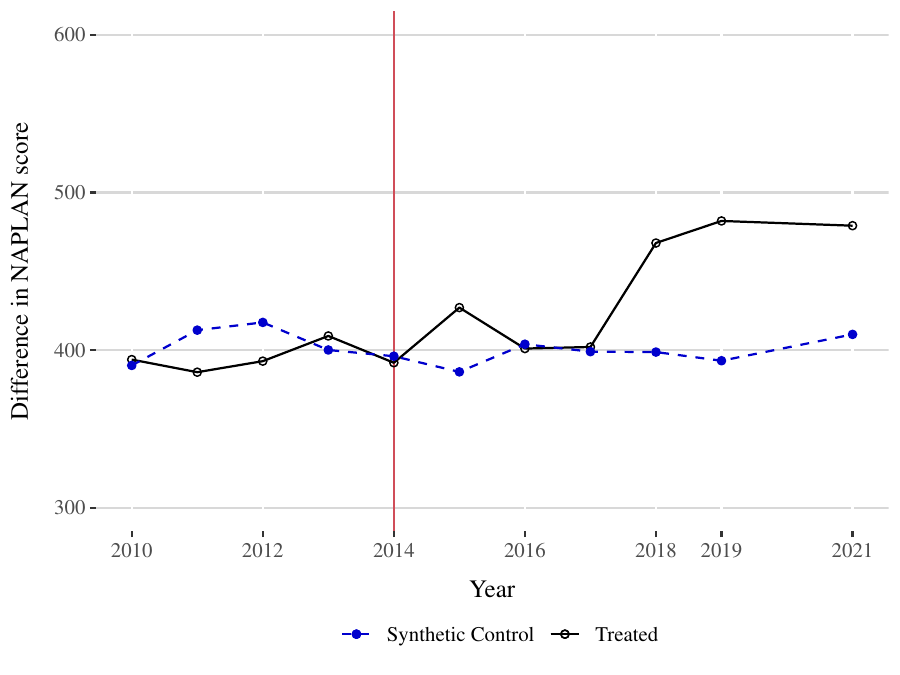}
        \label{fig:sc_num_y3}
    \end{subfigure}
    \begin{subfigure}{.5\textwidth}
        \centering
        \caption{Numeracy Year-5}
        \includegraphics[width=\linewidth]{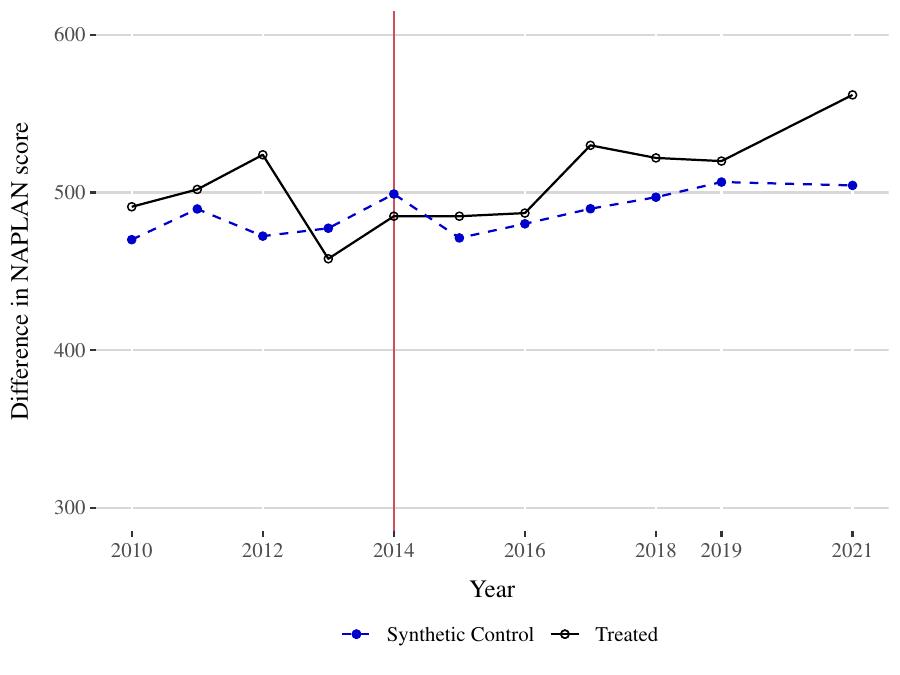}
        \label{fig:sc_num_y5}
    \end{subfigure}
    \newline
    \begin{subfigure}{.5\textwidth}
        \centering
        \caption{Reading Year-3}
        \includegraphics[width=\textwidth]{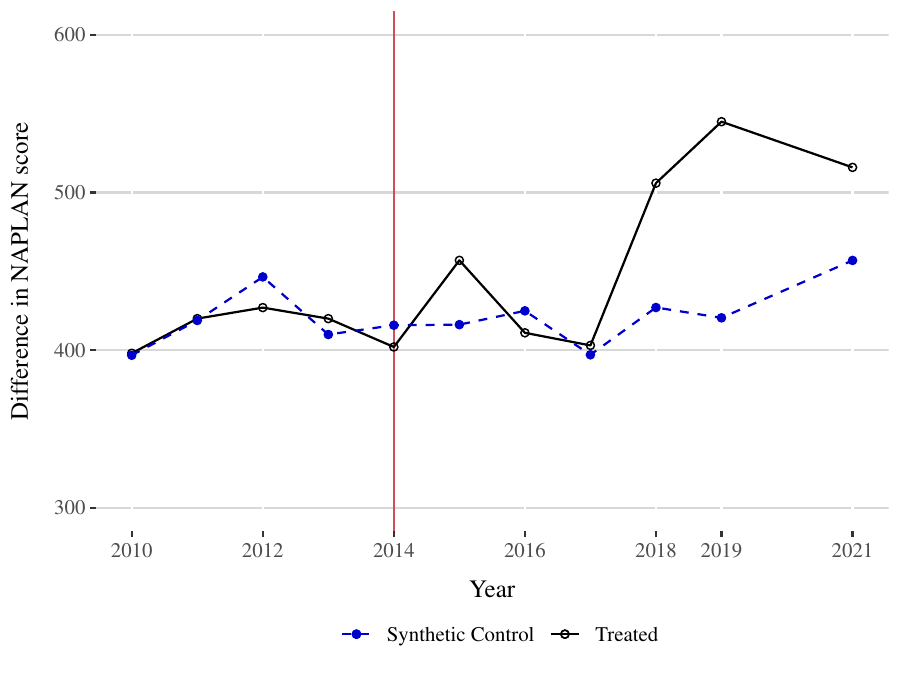}
        \label{fig:sc_read_y3}
    \end{subfigure}
    \begin{subfigure}{.5\textwidth}
        \centering
        \caption{Reading Year-5}
        \includegraphics[width=\linewidth]{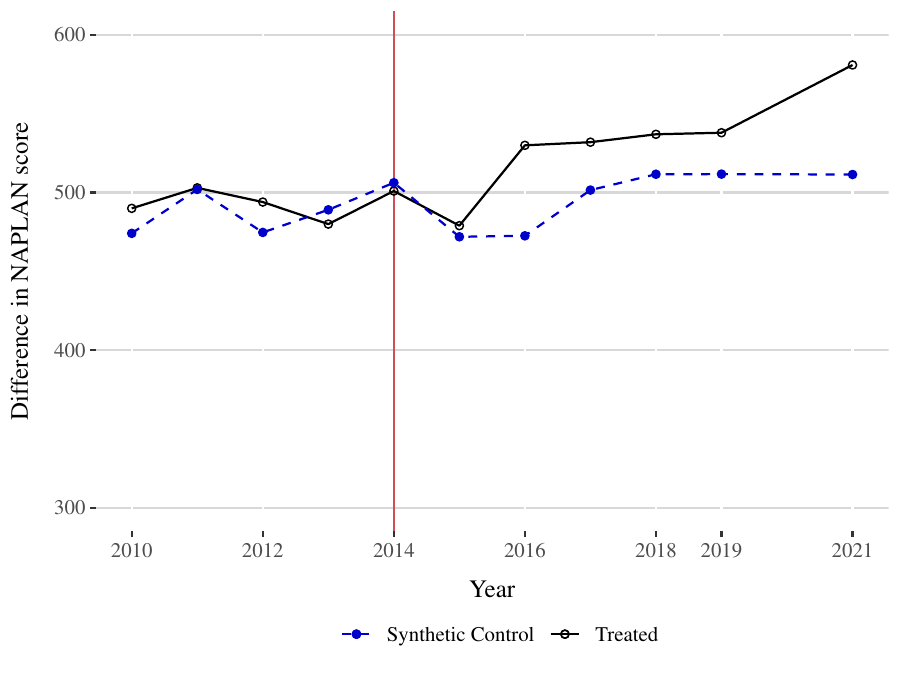}
        \label{fig:sc_read_y5}
    \end{subfigure}
    \caption{Effect of Explicit Instruction on NAPLAN scores}
    \label{fig:sc_num_read}
\end{figure}

\begin{figure}[h!]
    \begin{subfigure}{.5\textwidth}
        \centering
        \caption{\citeauthor{abadie10} \cite{abadie10} without covariates}
        \includegraphics[width=\linewidth]{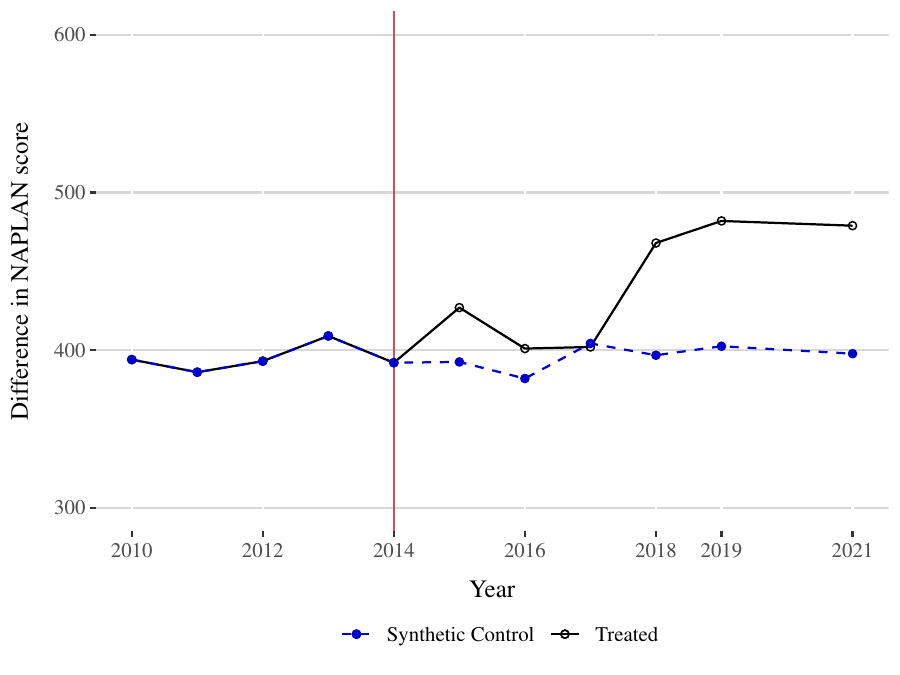}
        \label{fig:numeracy_Y3_nocov}
    \end{subfigure}
    \begin{subfigure}{.5\textwidth}
        \centering
        \caption{\citeauthor{ferman21} \cite{ferman21}}
        \includegraphics[width=\linewidth]{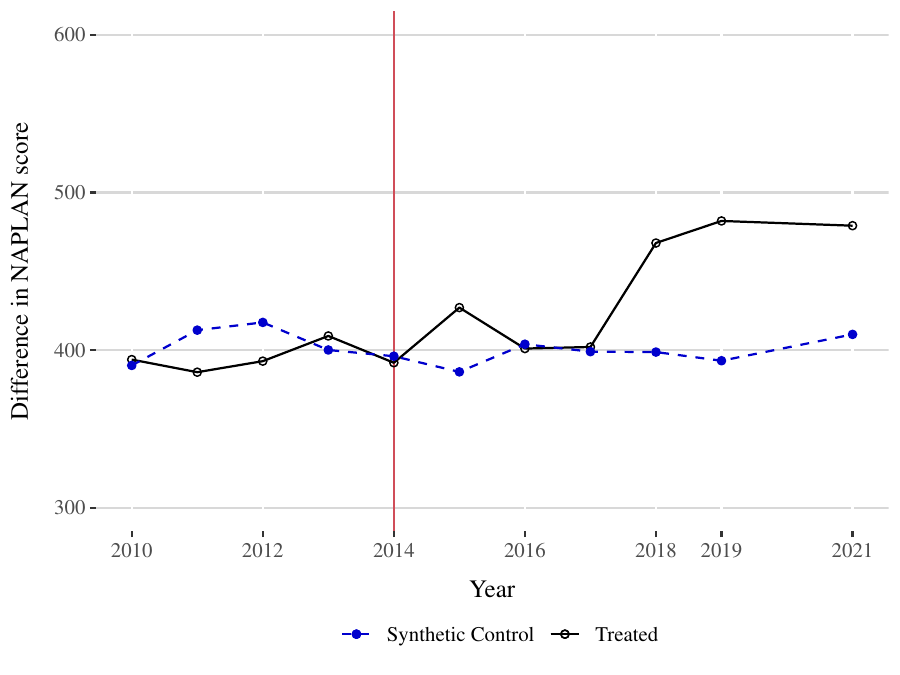}
        \label{fig:numeracy_Y3_ferman21}
    \end{subfigure}
    \newline
    \begin{subfigure}{.5\textwidth}
        \centering
        \caption{\citeauthor{hsiao12} \cite{hsiao12}}
        \includegraphics[width=\linewidth]{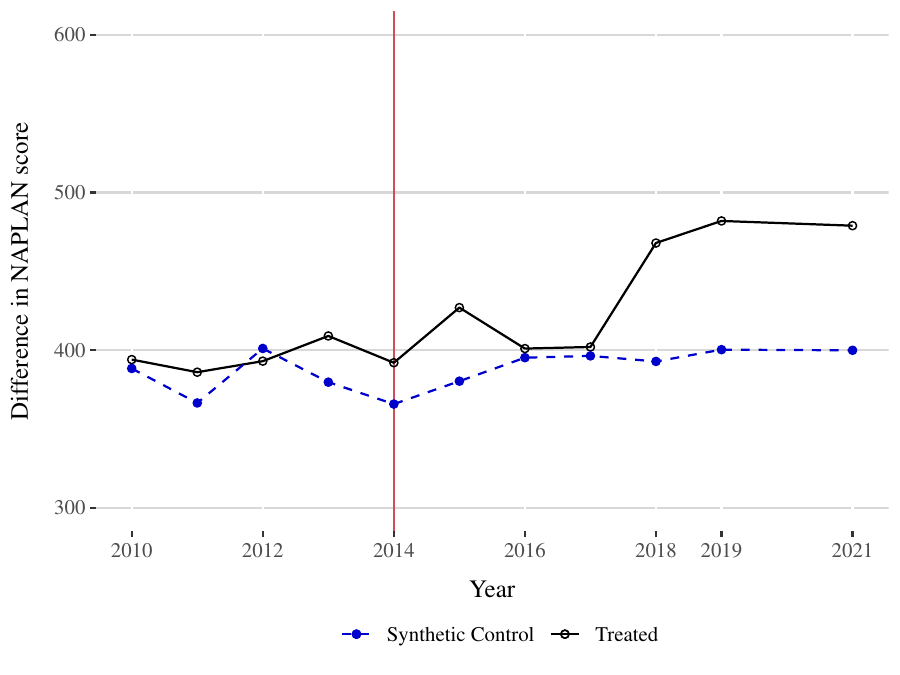}
        \label{fig:numeracy_Y3_hsiao12}
    \end{subfigure}
    \begin{subfigure}{.5\textwidth}
        \centering
        \caption{\citeauthor{chern21} \cite{chern21}}
        \includegraphics[width=\linewidth]{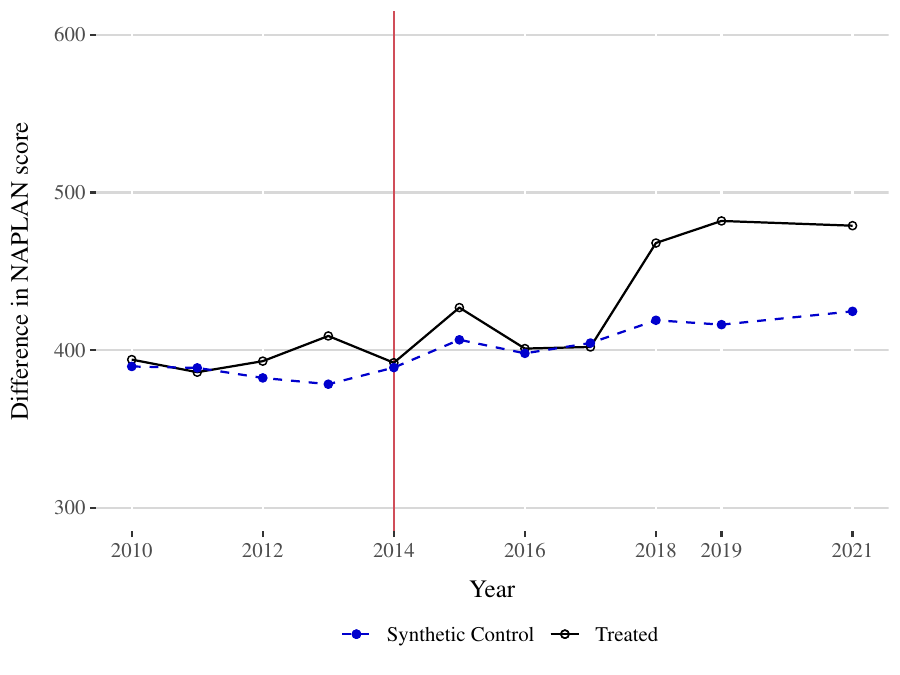}
        \label{fig:numeracy_Y3_chernoz21}
    \end{subfigure}
    \caption{Sensitivity analysis: Numeracy Year-3}
    \label{fig:numeracy_Y3_sensitivity}
\end{figure}

The results of our analysis are summarized in Figure \ref{fig:sc_num_read}. 
The patterns are similar across topics and are stable
across different estimators, as shown in Figures
\ref{fig:numeracy_Y3_sensitivity}, \ref{fig:numeracy_Y5_sensitivity},
\ref{fig:reading_Y3_sensitivity} and \ref{fig:reading_Y5_sensitivity} in
the Appendix.

The scores of students in Year 3 are positively affected starting three
years after Explicit Instruction was introduced at Charlestown
South. This is consistent with what was reported by the school's principal when interviewed, who stated that it took ``two or three years to really get a handle on it [the approach]''\footnote{The interview transcript is available upon request}.

In light of this, we focus on the effect strictly after three years from the
introduction of Explicit Instruction and find an average effect of 75.66 points (or 2.72 standard deviations) on Year-3 Numeracy scores and of 87.50 NAPLAN points (or 2.55 s.d) on Reading scores (see Figures \ref{fig:sc_num_y3} and \ref{fig:sc_read_y3}). Even if we want to calculate the treatment effect on Year-3 NAPLAN scores starting from the year when
Explicit Instruction was first introduced, its impact remains as large as 44.66 points (or 1.60 standard deviations) in Numeracy and 49.22 points (or 1.44 s.d.) in Reading.

The patterns in NAPLAN scores are also similar for Year-5 students across topics (see Figures \ref{fig:sc_num_y5} and \ref{fig:sc_read_y5}). Here, the effect of Explicit Instruction seems to start sooner---one or two years after the approach is introduced---but the magnitude is substantially smaller. If we continue to focus on the average treatment effect strictly three years after the treatment date, we find effects of less than half in size compared to Year-3 impacts\textemdash 31.90 points (or 1.14 s.d.) in Numeracy and 40.38 points (or 1.31 s.d.) in Reading. As above, even if we start calculating the effects for Year-5 students immediately after the pedagogy was officially introduced, we find large effects of 26.10 points (or 0.93 s.d.) in Numeracy and 36 points (or 1.17 s.d.) in Reading.

\begin{figure}[h!]
    \begin{subfigure}{.5\textwidth}
        \centering
        \caption{Numeracy Year-3}
        \includegraphics[width=\textwidth]{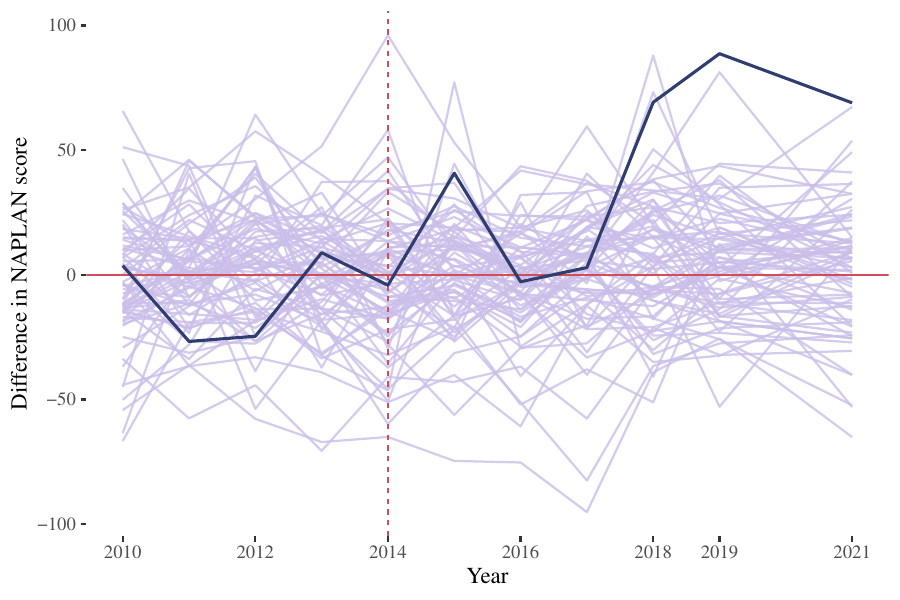}
        \label{fig:plac_num_y3}
    \end{subfigure}
    \begin{subfigure}{.5\textwidth}
        \centering
        \caption{Numeracy Year-5}
        \includegraphics[width=\linewidth]{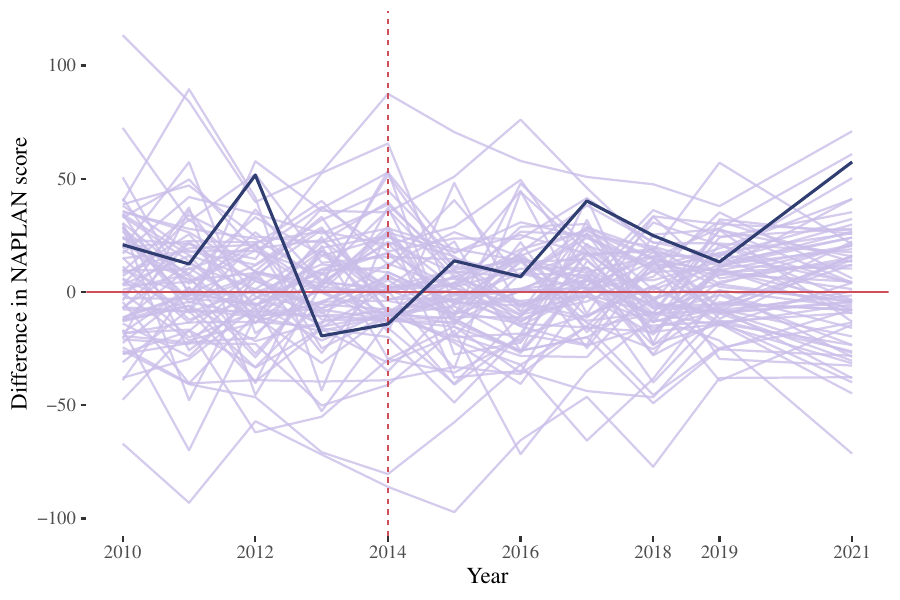}
        \label{fig:plac_num_y5}
    \end{subfigure}
    \newline
    \begin{subfigure}{.5\textwidth}
        \centering
        \caption{Reading Year-3}
        \includegraphics[width=\textwidth]{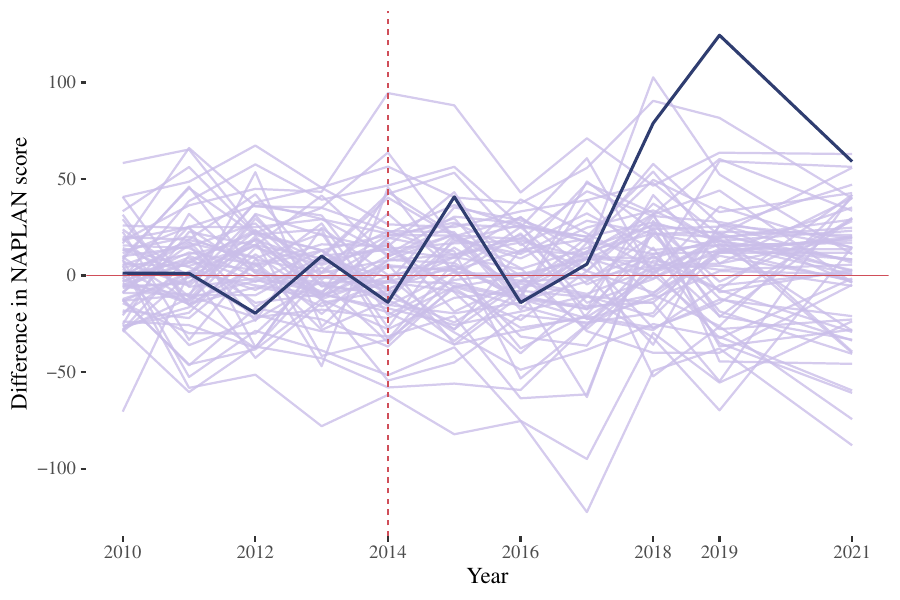}

        \label{fig:plac_read_y3}
    \end{subfigure}
    \begin{subfigure}{.5\textwidth}
        \centering
        \caption{Reading Year-5}
        \includegraphics[width=\linewidth]{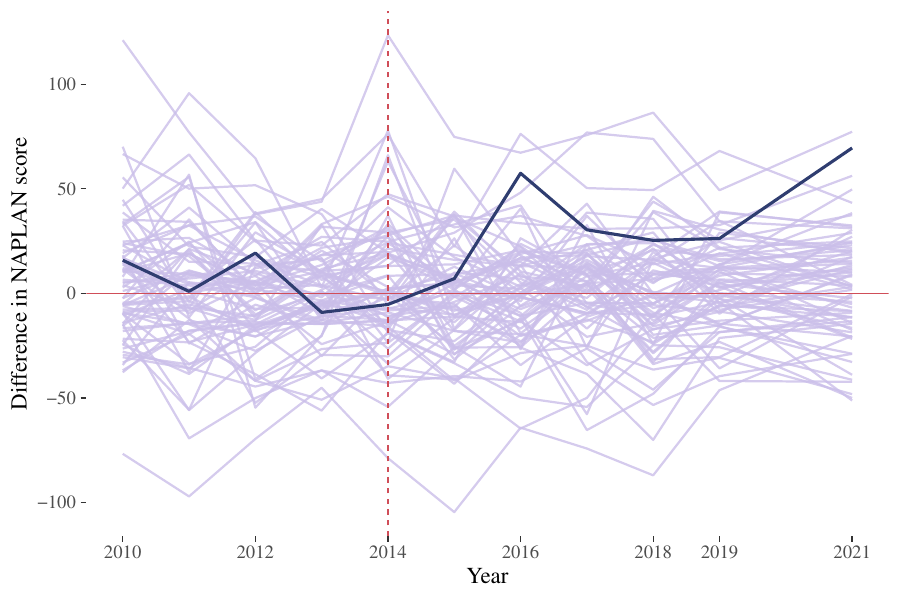}

        \label{fig:plac_read_y5}
    \end{subfigure}
    \caption{
        Placebo analysis on the effect of Explicit Pedagogy on NAPLAN scores
    }
    \label{fig:plac}
    {\vspace{2mm} \footnotesize \textit{Notes:} The figure visualizes the
        placebo test. For each school in the sample, it plots the gap between the path of the treated school and the path of the synthetic school\textemdash as estimated by the synthetic control estimator. The gap of Charlestown South is in blue while the placebo gaps are in purple.\par}    
\end{figure}

Moreover, when we run a placebo test for each of our four outcomes
and construct the respective exact p-values, we are unable to reject the
hypothesis that Explicit Instruction has no effect whatsoever
on the Year-5 Numeracy NAPLAN scores in any school at any time.
Conversely, we reject the equivalent hypotheses for the Year-3 Numeracy scores and the Reading scores in Year-3 and 5 (see Table \ref{tab:pval}).
A visual representation of the placebo test is provided in Figure
\ref{fig:plac}.

\begin{table}[h]
    \centering
    \begin{threeparttable}
        \caption{Exact p-values for each topic-year}
        \centering
        \begin{tabularx}{\textwidth}{XXX}
            \toprule
                     & Year 3 & Year 5 \\
            \toprule
            Numeracy & 0.027  & 0.347  \\
            Reading  & 0.015  & 0.013  \\
            \bottomrule
        \end{tabularx}
        \begin{tablenotes}
            \footnotesize
            \item \textbf{Note.} The hypothesis tested is
            $H_0: Y_{i, t}(1) = Y_{i, t}(0)$ for each
            school $i \in {1,..., N}$ and time period $t \in {1,..., T}$
            The significance level is set at $\alpha=0.1$. We cannot reject $H_0$
            only for the pedagogy impact estimates associated with Year-5
            Numeracy scores.
        \end{tablenotes}
        \label{tab:pval}
    \end{threeparttable}
\end{table}

To further test the sensitivity of our estimates, we take a leave-one-out approach. Specifically, we select the top 8 donor schools by weight and run our main specification 8 times, each time excluding one of these schools from the donor pool. As shown in Table \ref{tab:LOO} and Figures \ref{fig:loo1} and \ref{fig:loo2} in the appendix, the results are remarkably similar to our main estimates\textemdash both when taking each of the 8 sets of estimates individually (Figures \ref{fig:loo1} and \ref{fig:loo2}) and when looking at their aggregate average treatment effect (Table \ref{tab:LOO}). This is evidence that no individual donor school is driving the results.

Finally, our results are neither sensitive to the exclusion of covariates 
nor to the choice of synthetic control estimator. 
We show this in the Appendix, Figures 
\ref{fig:numeracy_Y3_sensitivity}, \ref{fig:numeracy_Y5_sensitivity},
\ref{fig:reading_Y3_sensitivity} and \ref{fig:reading_Y5_sensitivity}.
In panel (a) of each figure, we plot the results using the 
\citeauthor{abadie10} \cite{abadie10} estimator without using any covariates. 
In the (b) panels, we use the \citeauthor{ferman21} \cite{ferman21} estimator, which is 
a modification of \citeauthor{abadie10} \cite{abadie10} where the pretreatment average 
is calculated for all units and the data demeaned.
In the (c) panels, we use the \citeauthor{hsiao12} \cite{hsiao12} estimator, where 
weight vectors $\mathbf{W}$ is estimated via ordinary least 
squares, and are hence not constrained to sum up to one (see appendix for more methodological context).
In the (d) panels,  we use the \citeauthor{chern21} \cite{chern21} estimator, 
where weight vector $\mathbf{W}$ is estimated via 
constrained Lasso, hence relaxing the constraint on 
$\mathbf{W}$ compared to \citeauthor{abadie10} \cite{abadie10}\footnote{The \citeauthor{chern21} \cite{chern21} estimator allows $\mathbf{W}$ to 
be any vector with bounded $\ell_{1}$-norm.}.
Our results remain remarkably consistent.

\section{Discussion}
As mentioned, our estimated treatment effects range between 0.9 and 2.7 s.d. in size and apply to a median-performing school. According to \citeauthor{kraft2020} \cite{kraft2020}'s review of 747 RCTs, this places our intervention in the 99th percentile of education interventions by effect size. For context, a median intervention has an impact of 0.10 s.d. and a 90th percentile intervention of 0.40 s.d. \citep{kraft2020}.

Given the impact size alone of our intervention, there might be concerns that it is the leadership of the principal, rather than Explicit Instruction itself driving this radical change. This is only correct in the sense that it was the principal who identified that peer-modelling a top explicit-instruction school could turn his school into a high-performing one. I present evidence supporting this view below.

First, the principal started working at Charlestown South in 2012, while the change to Explicit Instruction started in 2014. Leadership and other principal-specific effects would have likely materialised before 2014 if they were meaningful. 
Second\textemdash when interviewed for another research project\textemdash the principal did not report any other intervention being introduced during his tenure. Third, word of his approach of peer-modelling a successful explicit-instruction school spread among other schools, which started sending staff to model Charlestown South. Indeed, the demand for learning Charlestown South approach directly became such that in 2019 the principal set up a termly ``EI Open Day'' dedicated to showing to the principals and staff from other schools the Explicit Instruction approach in action (\citealp{EducationHQ}; \citealp{CS_plan}; and see Facebook post \citealp{FB_CSPS}, for an example, also reported in Figure \ref{fig:fb} in the appendix). 

Not only the teaching community believes this specific implementation of Explicit Instruction is ideal and replicable\textemdash state governments believe it too. Indeed, in 2024, the Victorian state government tasked at LaTrobe University \citep{LaTrobe_MomentumSchools_2024} to link successful Explicit Instruction schools with schools that have not adopted this pedagogy yet. This will support the state-wide transition to structured phonics and explicit instruction among students from the first year of compulsory schooling (\textit{Prep}) to grade two. 

Finally, if we had to identify one key channel through which Explicit Instruction had an impact, it would be literacy, math and writing warm-ups. Indeed, both the high-performing Melbourne school and\textemdash after switching to Explicit Instruction\textemdash Charlestown South credit the warm-ups as the main driver of student performance, thanks to their ability to move information acquired in the previous lesson from the short-term to the long-term memory. 
The ``warm-up'' (or ``Daily Review'') is a 5-20-minute routine scheduled at the start of a lesson that rehearses past literacy, mathematics and writing content, and checks prerequisite skills, hence preparing students to learn new material  \citep{NSWExplicitTeaching2025,AISNSWDailyReview2025}. 
While the exact design of the warm-up depends on the grade of the students, during literacy warm-ups students might engage in quick exercises like revisiting letter sounds, practicing common words, or constructing simple sentences.
In writing warm-ups, teachers may review sentence structures, practice punctuation, or brainstorm vocabulary. These activities are aimed at strengthening core reading and writing skills, and prepare students for more complex writing tasks.
In mathematics warm-ups, teachers may prompt students with rapid questions on number facts, counting sequences, or basic calculations. This helps solidify numerical understanding and mental math skills.
These warm-ups are interactive, often involving the whole class, making learning more effective and engaging \citep[see][for detailed recent materials on this topic]{NSWLiteracyStrategies2025,LiteracyHubDailyReview2025}. 

\paragraph{Concluding remarks.} We analyze the effect of correctly implementing Explicit Instruction on learning outcomes, as measured by standardized test scores. We focus on the case of primary public school Charlestown South, in New South Wales. We find large effects in both Numeracy and Reading, particularly on Year-3 students. Our sensitivity analysis provides evidence that the results are not an artifact either of our preferred method or of our preferred model specification. 

Our results identify an effective (and cost-effective) way to introduce Explicit Instruction\textemdash peer modelling. 
It consists in having the teaching staff of a target school sitting-in during the classes of a high-performing explicit-instruction school for as little as three days. This is enough to generate large and persistent performance gains in standardised tests. Because our target school was a median performer at baseline, the impact of this intervention on low-performing schools might be even larger.

\section{Materials and Methods \label{sec:method}}
To create a synthetic Charlestown South, first, we need to select a pool of schools that we judge as being similar to Charletown South, but were not subject to the same intervention\textemdash here, learning the Explicit Instruction pedagogy by peer-modelling a successful Explicit Instruction school. Among all primary public NSW schools, we selected schools that have some key features that are identical to Charlestown South. We exact-match these features as we see them as too important to be matched using the minimum-distance approach embedded in SCM (see the next paragraph).
They have the same level of remoteness \citep{asgs}, ``Major Cities of Australia''; they are co-educational; they cover Kindergarten to Year 6, and their student body comprises between 10-15\% of students with a language background other than English (Charlestown South's is 12\%). We are left with 108 schools which, in the language of synthetic control, are referred to as \textit{donor schools}. 

Second, we (minimum-distance) match donor schools to Charlestown South against the following variables: share of attendance, full-time equivalent enrollments, share of female students, mean taxable income at the school's postcode, ICSEA score, average class size, language background other than English, year in which the first teacher was hired, and radial distance from Charlestown South.

Third, we set the year of ``treatment'' at 2014, as this is the year when Charlestown South introduced Explicit Instruction. Hence, from 2014 onwards
we consider Charlestown South as ``treated'', while before it was ``untreated'', while the donor schools are always considered untreated.

Fourth, we construct a synthetic control. This involves creating a ``synthetic'' version of Charlestown South as similar as possible as the real Charlestown South. Since we cannot observe Charlestown South had it not been treated in 2014, we use its synthetic version to approximate this counterfactual scenario. This allows us to estimate the treatment effect of implementing Explicit Instruction via peer-modelling over time\textemdash by comparing Charlestown South after treatment
with its synthetic version. The synthetic control is built by
finding a ``mix'' (the convex hull) of donor schools such that associated
donor school weights (i) leading to pre-treatment paths of NAPLAN scores ``most similar'' to Charlestown South's and (ii) produce school features of synthetic Charlestown South that are ``most similar'' to those of the real one.

\bibliographystyle{unsrtnat}
\bibliography{report_charl_biblio}

\pagebreak

\section{Appendix}

\subsection{Estimation}
Formally, we observe the outcome of interest $Y_{i t}$\textemdash the NAPLAN
scores\textemdash for each school $i$ and year $t$. We also observe $K$
covariates (or features) per each school, $X_{1 i}, \ldots, X_{K i}$,
which are themselves unaffected by the intervention.
These covariates are selected for their capacity to affect NAPLAN scores.
If a covariate changes over time for a given school,
rather than being a fixed feature, we take the within-unit mean for that
covariate\textemdash following \citeauthor{abadie03} \cite{abadie03}.

In the matrix notation used below: (i)
the $K \times 1$ vectors $\mathbf{X}_{1}, \ldots, \mathbf{X}_{N}$ contains
the values of the covariates of units $i=1, \ldots, N$, respectively, and (ii)
the $K \times (N-1)$ matrix,
$\mathbf{X}_{0}=\left[\mathbf{X}_{1} \cdots \mathbf{X}_{N-1}\right]$,
collects the values of the covariates of the $N-1$ untreated (or \textit{donor}) schools.

We follow \citep*{abadie03} and \citeauthor{abadie10} \cite{abadie10} and choose weights
$\mathbf{W}^{*}=\left(w_{1}^{*}, \ldots, w_{N-1}^{*}\right)^{\prime}$
that minimize the Euclidean distance between the covariates of treated and
untreated schools in the pre-treatment periods, while restricting the SC weights
to being positive and summing up to one.

Formally,

\begin{align}
    \min_{\boldsymbol{W}}\left\|\mathbf{X}_{1}-\mathbf{X}_{0} \boldsymbol{W}\right\|
     & =\sqrt{\left(\mathbf{X}_{1}-\mathbf{X}_{0} \mathbf{W}\right)^{\prime} \mathbf{V}\left(\mathbf{X}_{1}-\mathbf{X}_{0} \mathbf{W}\right)}              \\
     & =\left(\sum _ { k = 1 } ^ { K } v _ { k } \left(X_{k N}-w_{1} X_{k 1}-\cdots\right.\right. \left.\left.-w_{N-1} X_{k N-1}\right)^{2}\right)^{1 / 2} \\
\end{align}

\begin{align}
     & \text{where } \mathcal{W}=\left\{\left(w_{1}, \ldots, w_{N-1}\right)^{\prime}\right. \text{subject to } & \hspace{5pt} w_{1} + \cdots + w_{N-1}=1, \\
     &                                                                                                         & \hspace{5pt} w_{i} \geq 0,               \\
     &                                                                                                         & \text{with }i = 1, \ldots, N-1. \}
\end{align}

$\mathbf{V}$ is a $(K \times K)$ symmetric and positive semidefinite matrix.
It is another matrix of weights. Indeed, one the one hand,
$\boldsymbol{W}$ regulates how much importance
each \textit{donor school} has in the newly created
synthetic Charlestown South\textemdash what share of
synthetic Charlestown South was donated by each donor school
(notice the $i$ index in $w_{i}$, which stands for school).

On the other hand, $\mathbf{V}$ sets the importance that each covariate
has in constructing synthetic Charlestown South.
To this end, $\mathbf{V}=\{v_1 + \cdots + v_K\}$ is chosen so that it best
predicts $Y_{Nt}(0)$, in the sense that it minimises the
Mean Squared Prediction Error (MSPE) between the pre-treatment outcome of the
treated school over time ($Y_{Nt}$, $t \leq T_{0}$)
and that of the synthetic control over time,
$\sum_{i=}^{N} w_{i}(\mathbf{V}) Y_{i t}$.

Formally, $\mathbf{V}^*$ solves

\begin{equation}
    \min_{\mathbf{V} \in \mathcal{V}} \sum_{t \in \mathcal{T}_{0}}\left(Y_{N t}-w_{1}(\boldsymbol{V}) Y_{1 t}-\cdots-w_{N-1}(\boldsymbol{V}) Y_{N-1 t}\right)^{2},
\end{equation}

for some set $\mathcal{T_0} \subseteq\left\{1,2, \ldots, T_{0}\right\}$ of
pre-treatment periods,
or equivalently, in matrix notation,
\begin{equation}
    \min_{\mathbf{V}\in\mathcal{V}}=\left(\mathbf{Z}_{1}-\mathbf{Z}_{0} \mathbf{W^{*}(\mathbf{V})}\right)^{\prime} \left(\mathbf{Z}_{1}-\mathbf{Z}_{0} \mathbf{W^{*}}(\mathbf{V})\right) \\
\end{equation}

where $\mathcal{V}$ is the set of all nonnegative diagonal $(K \times K)$
matrices, and $\mathbf{Z_1}$ and $\mathbf{Z_0}$ are a vector of pre-treatment
outcome values for the treated unit and a matrix of pre-treatment outcomes
for the donor units.
Finally, the final weights for the synthetic control are given by
$\mathbf{W}^{*}(\mathbf{V^{*}})$.

\subsection{Inference}
After estimating the effect of Explicit Instruction on NAPLAN scores in Charlestown South, we conduct a placebo test to check whether any effect we find is likely due to chance or not. To do so, we run the same SC analysis
on all possible permutations of treatment assignment (keeping the timing
of the treatment fixed at the original value). We run an additional 108 SC analyses, pretending in each that a different donor school was treated, rather than Charlestown South. We then use exact p-values to formally test whether
the effect sizes that we find for Charlestown South are "extreme" compared
to those found in the 108 placebo analyses. If they are extreme,
we take that as evidence that our results are not due to chance.

Formally, we calculate the exact p-value

\begin{equation}
    p:=\frac{\sum_{j=1}^{N} \mathbb{I}\left[R M S P E_{i} \geq R M S P E_{N}\right]}{N}
\end{equation}

where $\mathbb{I}[\mathbf{A}]$ is an indicator function and
$RMSPE_i$ is the ratio mean squared prediction error, defined as

\begin{equation}
    R M S P E_{i}:=\frac{\sum_{t=T_{0}+1}^{T}\left(Y_{i, t}-\widehat{Y_{i, t}(0)}\right)^{2} /\left(T-T_{0}\right)}{\sum_{t=1}^{T_{0}}\left(Y_{i, t}-\widehat{Y_{i, t}(0)}\right)^{2} / T_{0}}
\end{equation}

where the numerator is the post-treatment MSPE of school $i$ and the
denominator is the pre-treatment MSPE of that same school.
We use the above statistic to test the sharp null
\begin{equation}
    H_{0}: Y_{i, t}(1)=Y_{i, t}(0) \text { for each region } i \in\{1, \ldots, N\} \text { and time period } t \in\{1, \ldots, T\}
\end{equation}
which will or will not be rejected based on pre-set significance level
$\alpha=0.05$.

\subsection{Other results}

\begin{figure}[h!]
    \begin{subfigure}{.5\textwidth}
        \centering
        \caption{\citeauthor{abadie10} \cite{abadie10} without covariates}
        \includegraphics[width=\linewidth]{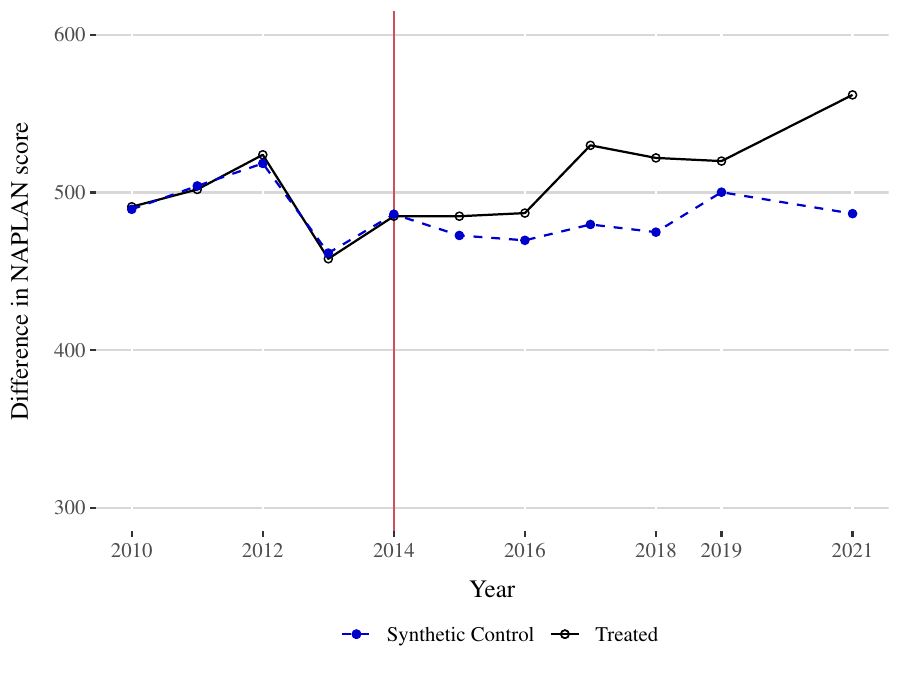}
        \label{fig:numeracy_Y5_nocov}
    \end{subfigure}
    \begin{subfigure}{.5\textwidth}
        \centering
        \caption{\citeauthor{ferman21} \cite{ferman21}}
        \includegraphics[width=\linewidth]{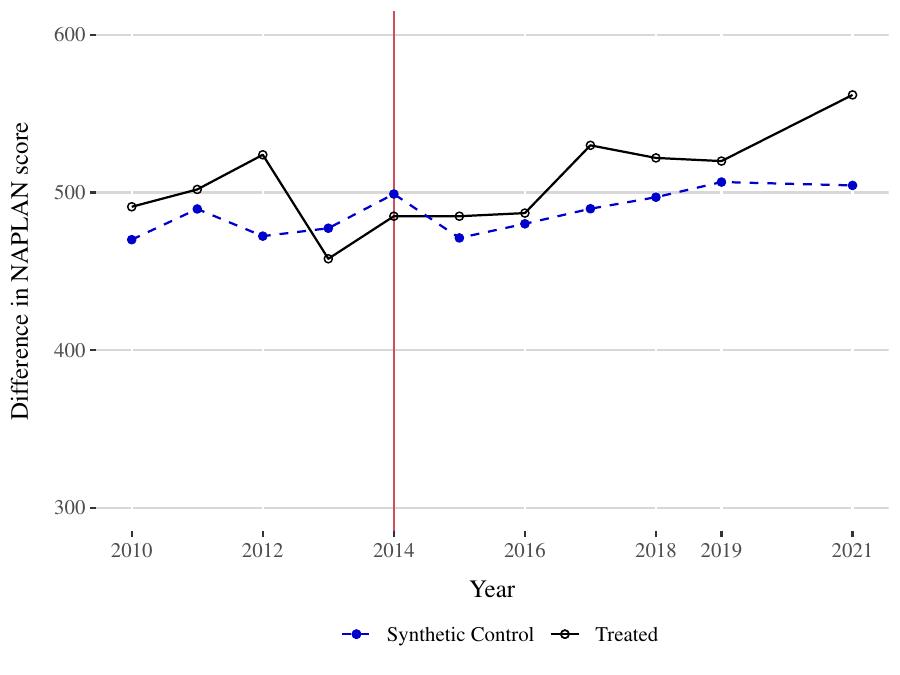}
        \label{fig:numeracy_Y5_ferman21}
    \end{subfigure}
    \newline
    \begin{subfigure}{.5\textwidth}
        \centering
        \caption{\citeauthor{hsiao12} \cite{hsiao12}}
        \includegraphics[width=\linewidth]{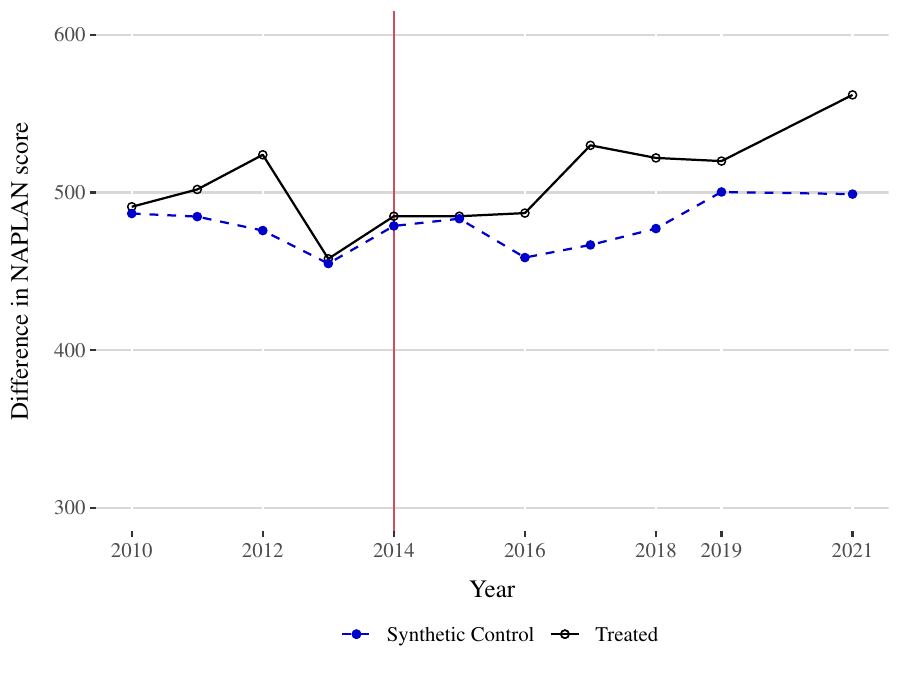}
        \label{fig:numeracy_Y5_hsiao12}
    \end{subfigure}
    \begin{subfigure}{.5\textwidth}
        \centering
        \caption{\citeauthor{chern21} \cite{chern21}}
        \includegraphics[width=\linewidth]{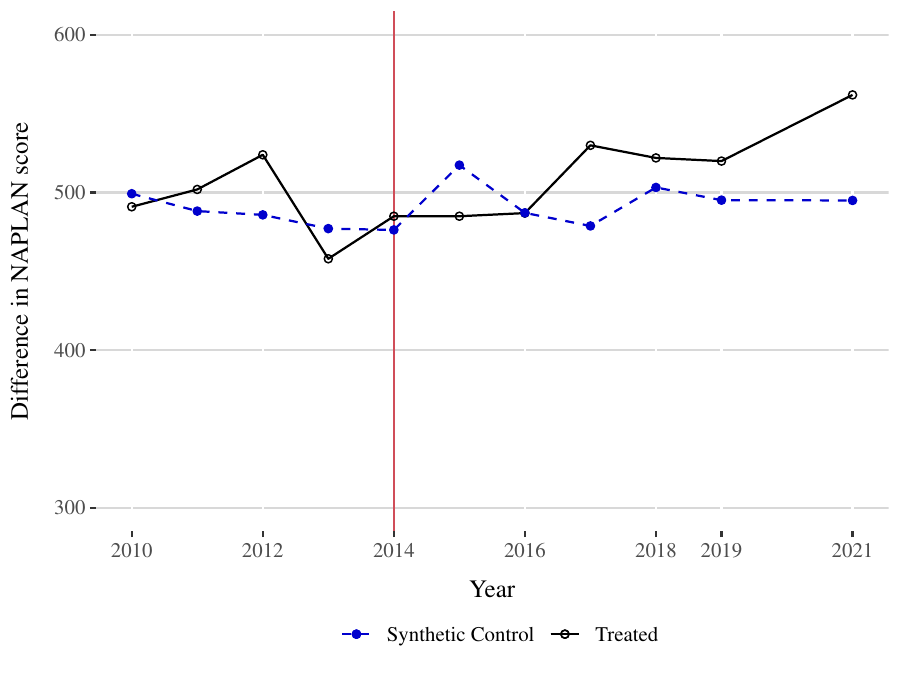}
        \label{fig:numeracy_Y5_chernoz21}
    \end{subfigure}
    \caption{Sensitivity analysis: Numeracy Year-5 }
    \label{fig:numeracy_Y5_sensitivity}
\end{figure}

\begin{figure}[h!]
    \begin{subfigure}{.5\textwidth}
        \centering
        \caption{\citeauthor{abadie10} \cite{abadie10} without covariates}
        \includegraphics[width=\linewidth]{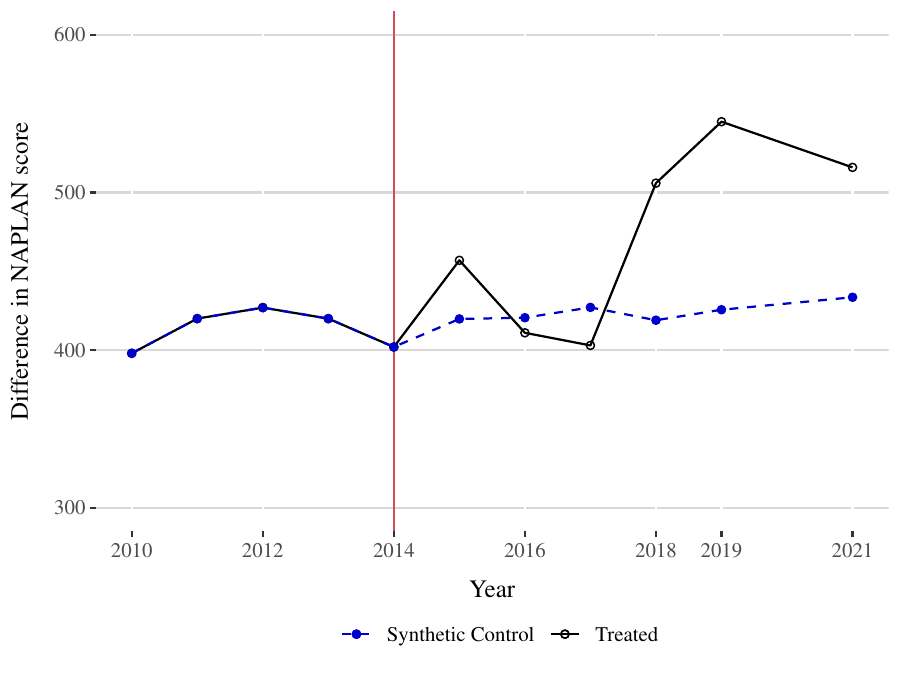}
        \label{fig:reading_Y3_nocov}
    \end{subfigure}
    \begin{subfigure}{.5\textwidth}
        \centering
        \caption{\citeauthor{ferman21} \cite{ferman21}}
        \includegraphics[width=\linewidth]{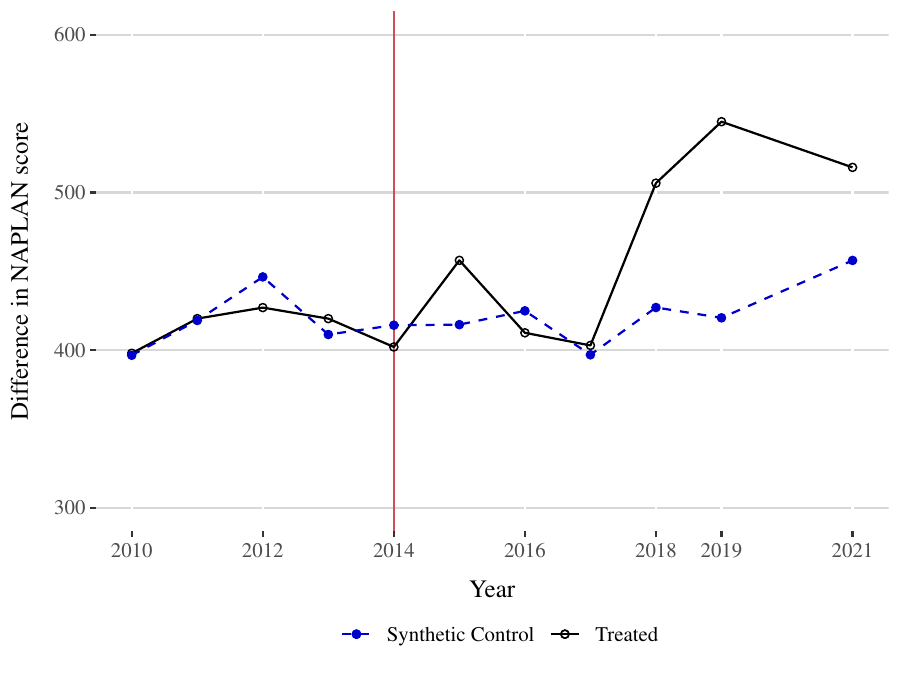}
        \label{fig:reading_Y3_ferman21}
    \end{subfigure}
    \newline
    \begin{subfigure}{.5\textwidth}
        \centering
        \caption{\citeauthor{hsiao12} \cite{hsiao12}}
        \includegraphics[width=\linewidth]{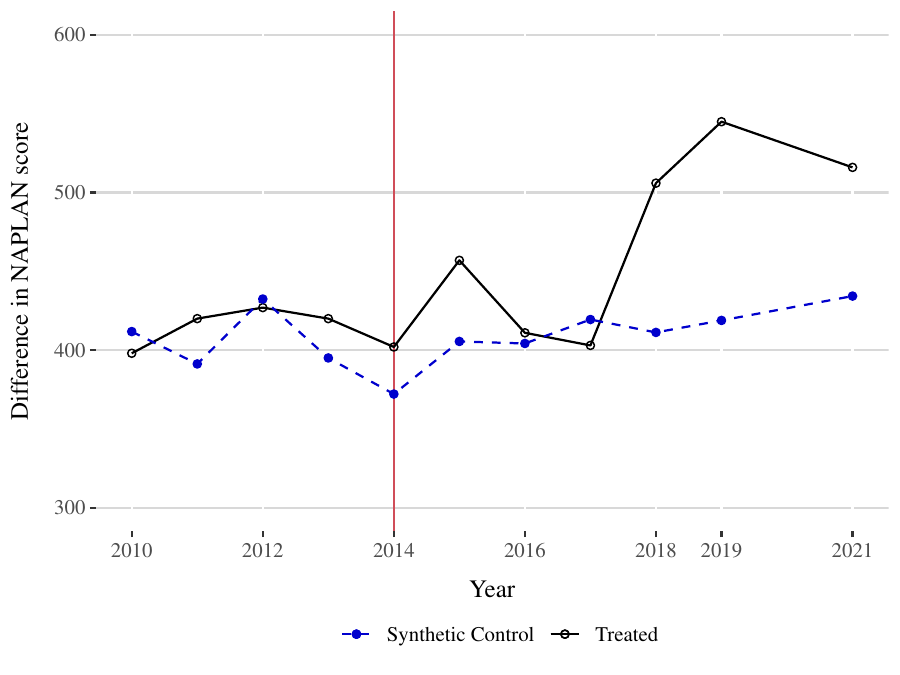}
        \label{fig:reading_Y3_hsiao12}
    \end{subfigure}
    \begin{subfigure}{.5\textwidth}
        \centering
        \caption{\citeauthor{chern21} \cite{chern21}}
        \includegraphics[width=\linewidth]{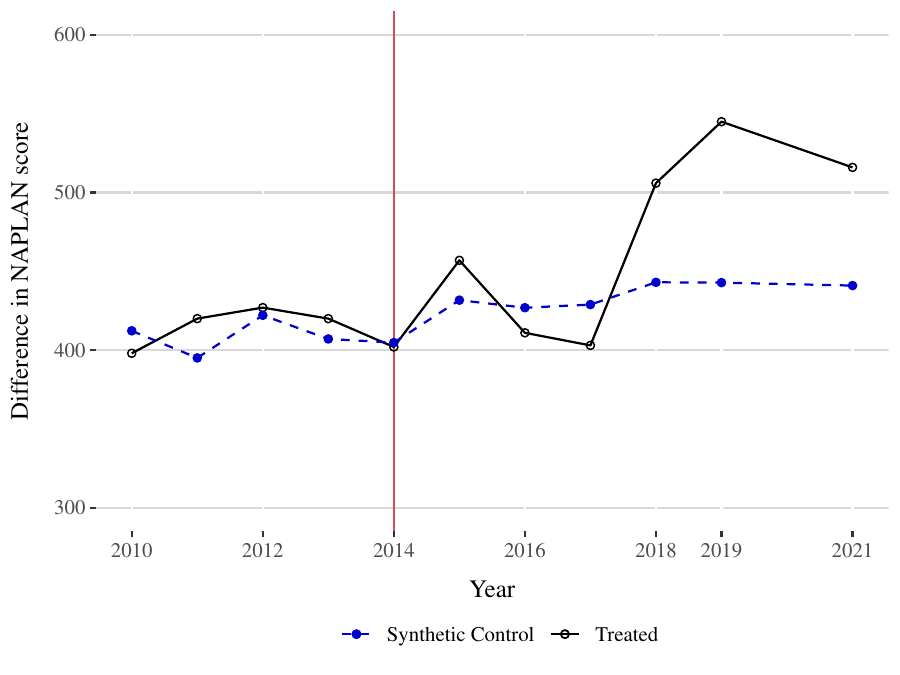}
        \label{fig:reading_Y3_chernoz21}
    \end{subfigure}
    \caption{Sensitivity analysis: Reading Year-3}
    \label{fig:reading_Y3_sensitivity}
\end{figure}

\begin{figure}[h!]
    \begin{subfigure}{.5\textwidth}
        \centering
        \caption{\citeauthor{abadie10} \cite{abadie10} without covariates}
        \includegraphics[width=\linewidth]{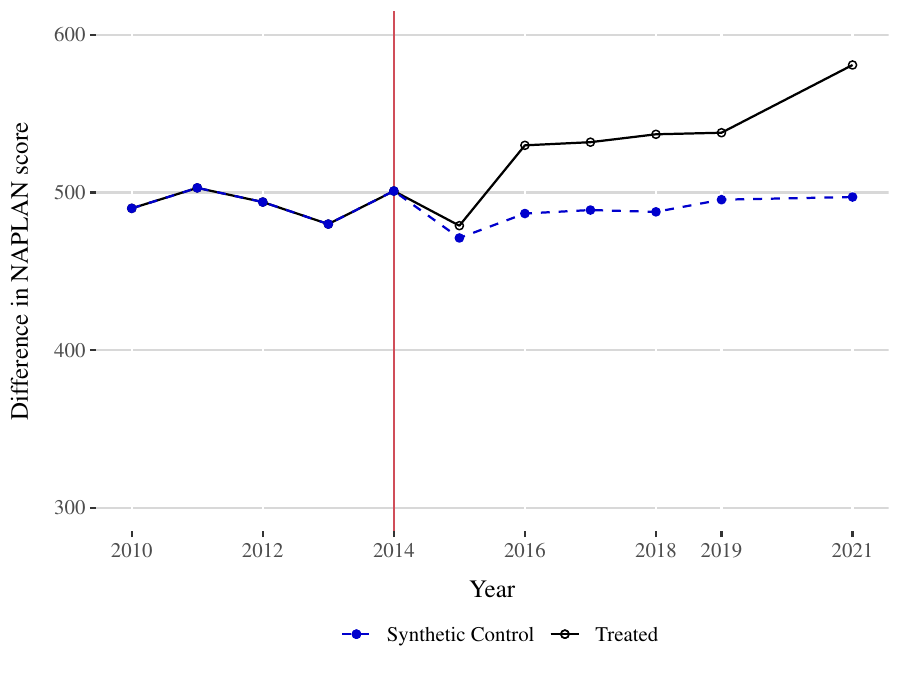}
        \label{fig:reading_Y5_nocov}
    \end{subfigure}
    \begin{subfigure}{.5\textwidth}
        \centering
        \caption{\citeauthor{ferman21} \cite{ferman21}}
        \includegraphics[width=\linewidth]{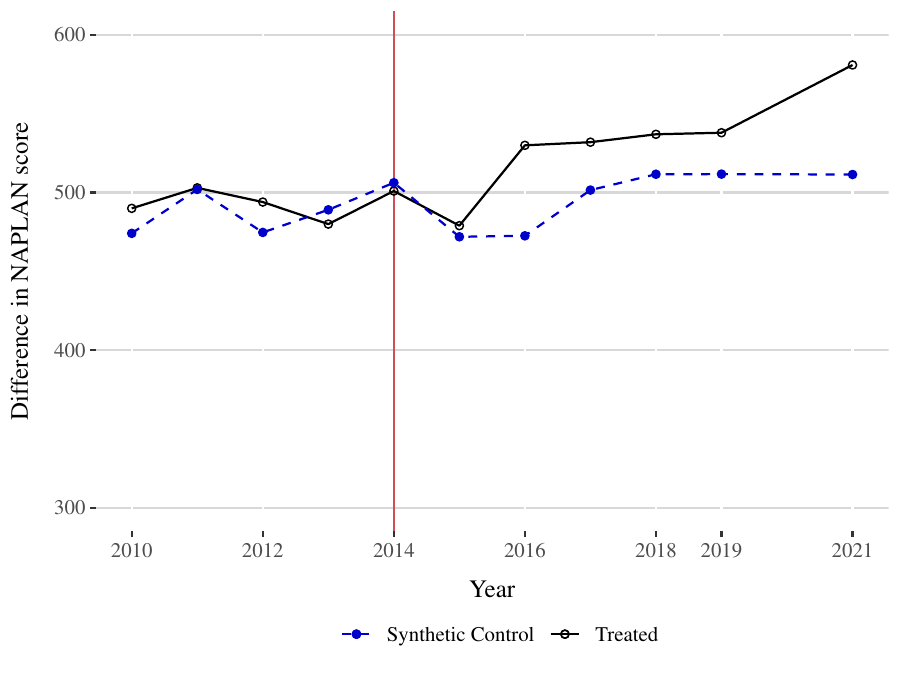}
        \label{fig:reading_Y5_ferman21}
    \end{subfigure}
    \newline
    \begin{subfigure}{.5\textwidth}
        \centering
        \caption{\citeauthor{hsiao12} \cite{hsiao12}}
        \includegraphics[width=\linewidth]{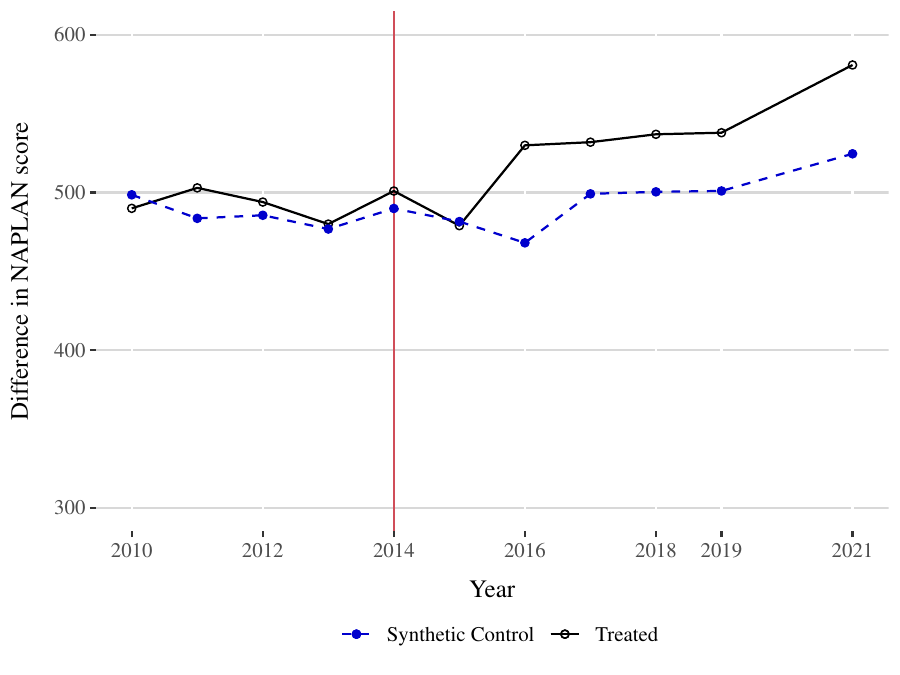}
        \label{fig:reading_Y5_hsiao12}
    \end{subfigure}
    \begin{subfigure}{.5\textwidth}
        \centering
        \caption{\citeauthor{chern21} \cite{chern21}}
        \includegraphics[width=\linewidth]{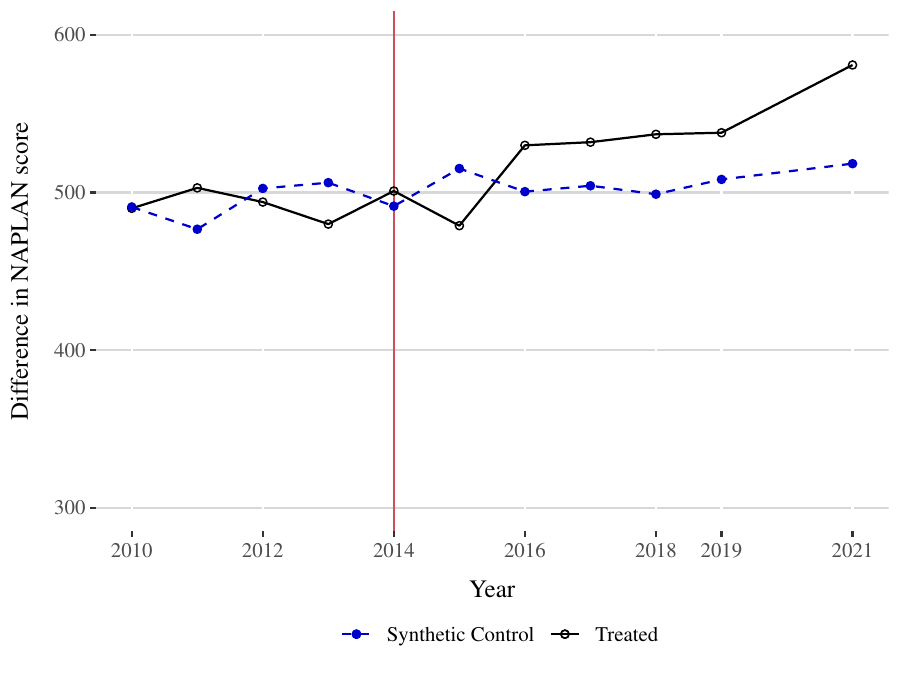}
        \label{fig:reading_Y5_chernoz21}
    \end{subfigure}
    \caption{Sensitivity analysis: Reading Year-5 }
    \label{fig:reading_Y5_sensitivity}
\end{figure}


\begin{figure}[htb]
    \centering
    \thispagestyle{empty}
    \begin{tikzpicture}
        \node (img1)  {\includegraphics[width=0.20\textwidth]{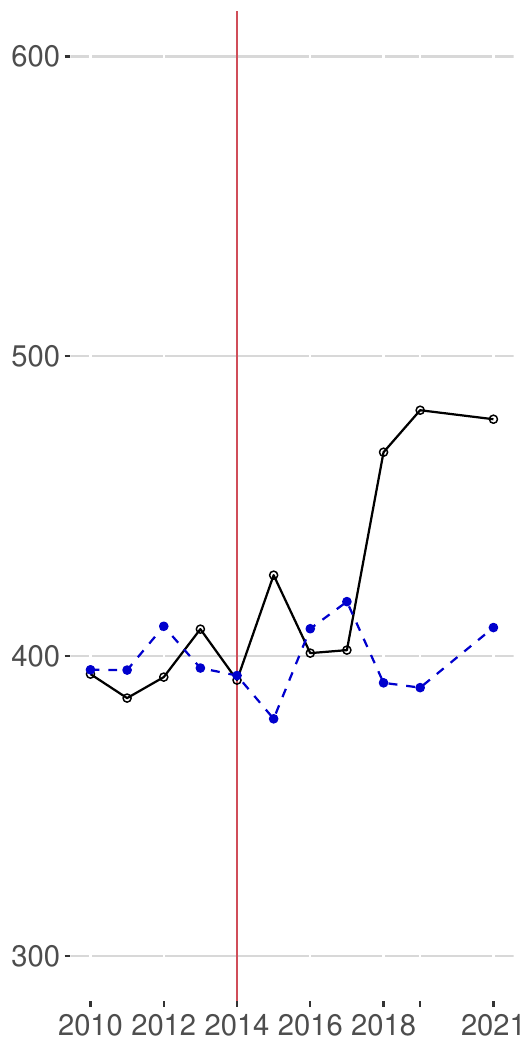}};
        \node[above=of img1, node distance=0cm, yshift=-1cm,font=\color{black}] {Numeracy Y-3};
        \node[left=of img1, node distance=0cm, xshift=-0.5cm, rotate=90, anchor=center,yshift=-1cm,font=\color{black}] {\small Naplan score};
        \node[left=of img1, node distance=0cm, rotate=90, anchor=center,yshift=0cm,font=\color{black}] {Charlestown (North)};

        \node[right=of img1, yshift=0cm, xshift=-1cm] (img2)  {\includegraphics[width=0.20\textwidth]{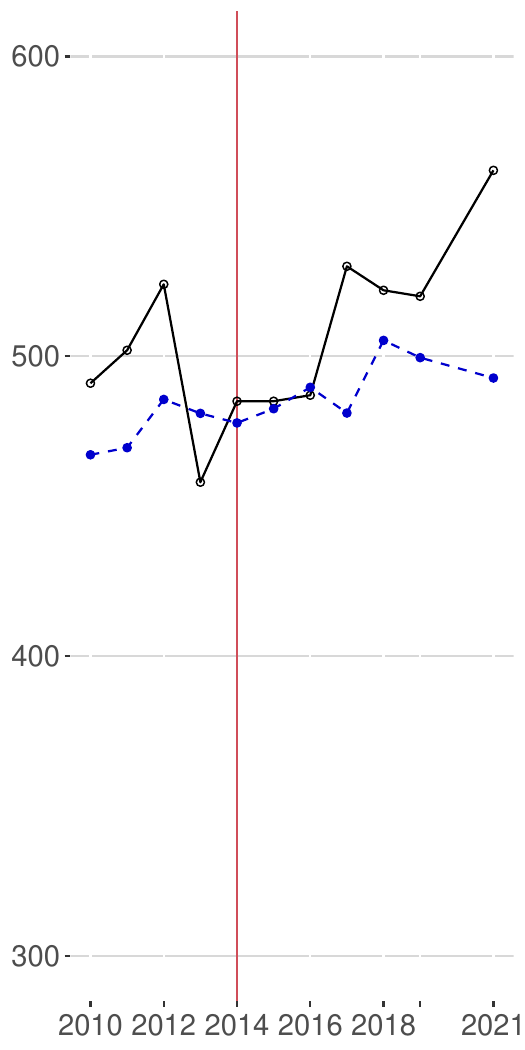}};
        \node[above=of img2, node distance=0cm, yshift=-1cm,font=\color{black}] {Numeracy Y-5};

        \node[right=of img2, yshift=0cm, xshift=-1cm] (img3)  {\includegraphics[width=0.20\textwidth]{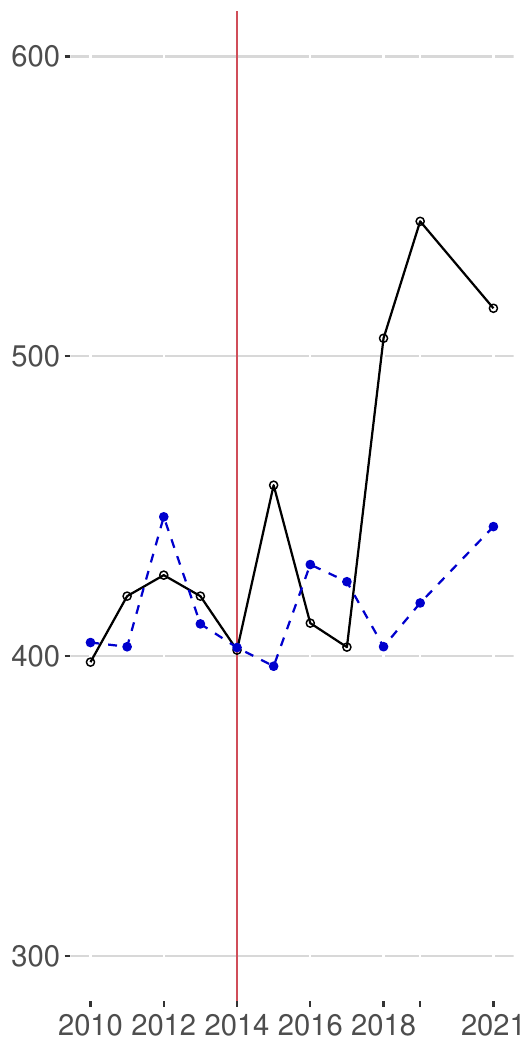}};
        \node[above=of img3, node distance=0cm, yshift=-1cm,font=\color{black}] {Reading Y-3};
        
        \node[right=of img3, yshift=0cm, xshift=-1cm] (img4)  {\includegraphics[width=0.20\textwidth]{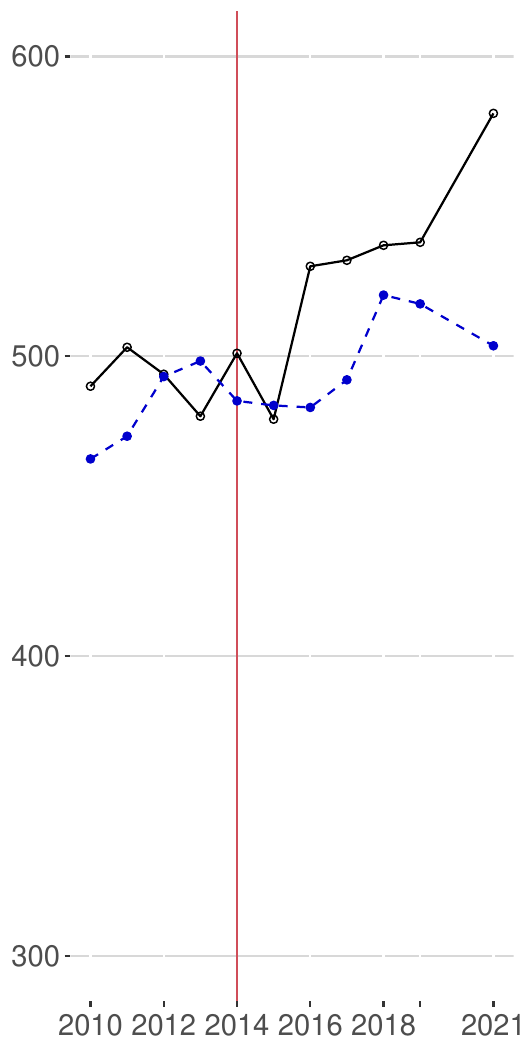}};
        \node[above=of img4, node distance=0cm, yshift=-1cm,font=\color{black}] {Reading Y-5};

        \node[below=of img1, yshift=1cm] (img5)  {\includegraphics[width=0.20\textwidth]{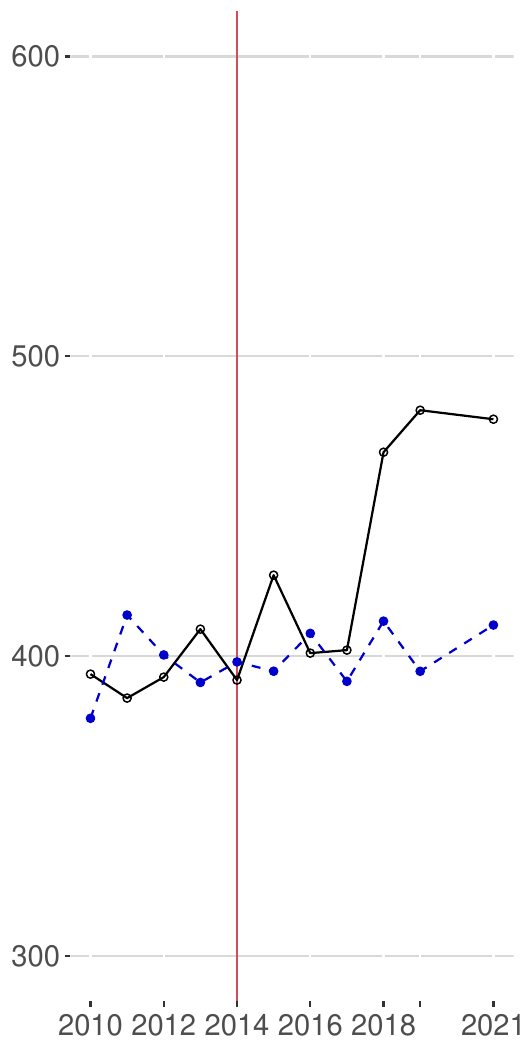}};
        \node[below=of img2, yshift=1cm] (img6)  {\includegraphics[width=0.20\textwidth]{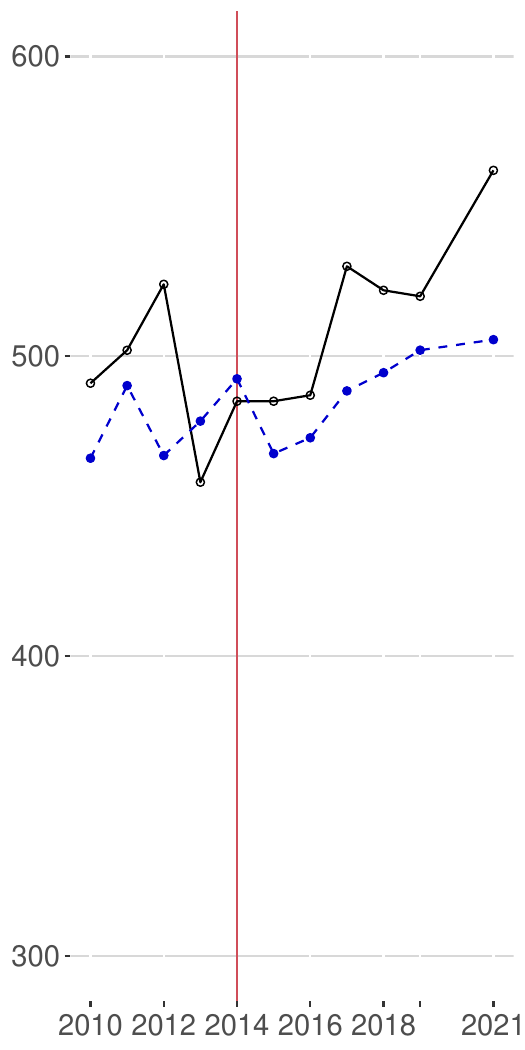}};
        \node[below=of img3, yshift=1cm] (img7)  {\includegraphics[width=0.20\textwidth]{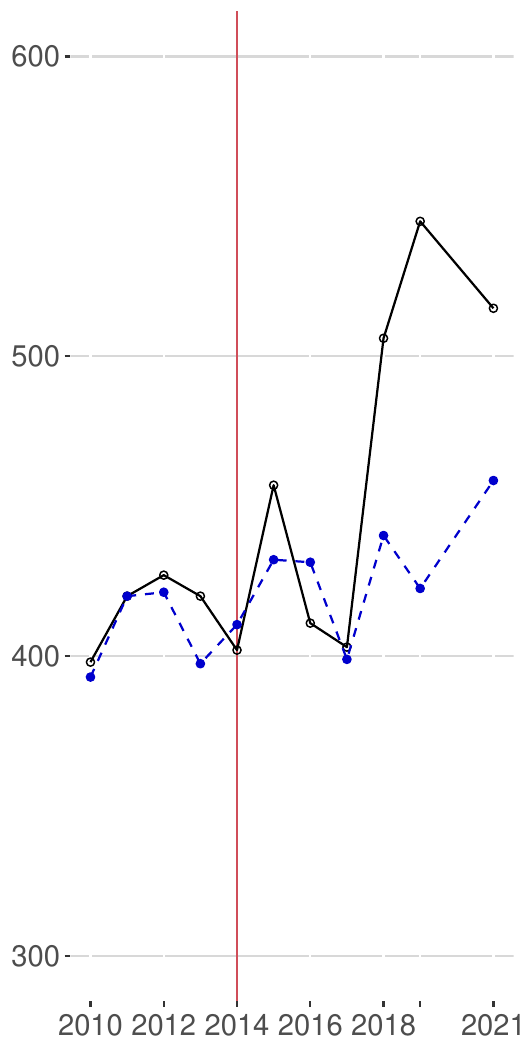}}; 
        \node[below=of img4, yshift=1cm] (img8)  {\includegraphics[width=0.20\textwidth]{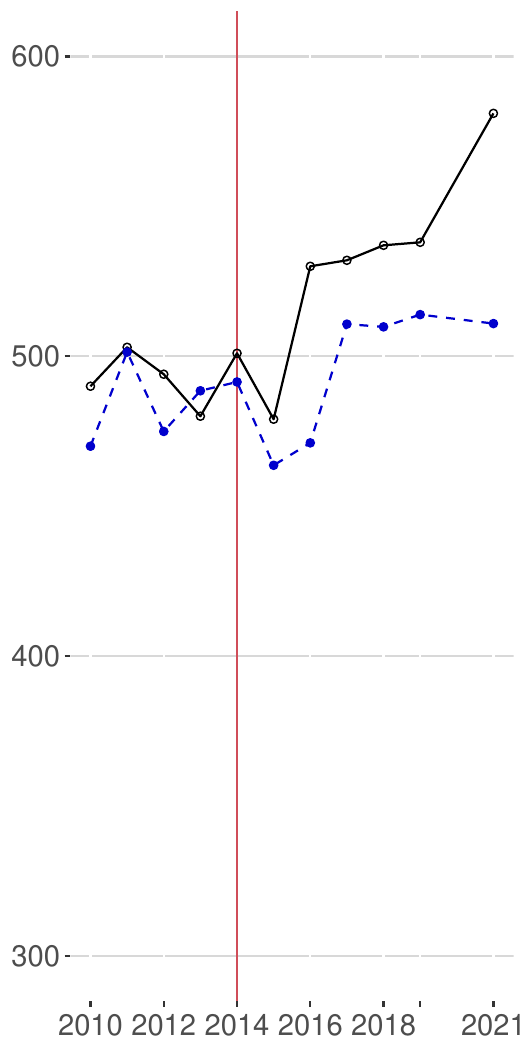}}; 

        \node[below=of img5, yshift=1cm] (img9)  {\includegraphics[width=0.20\textwidth]{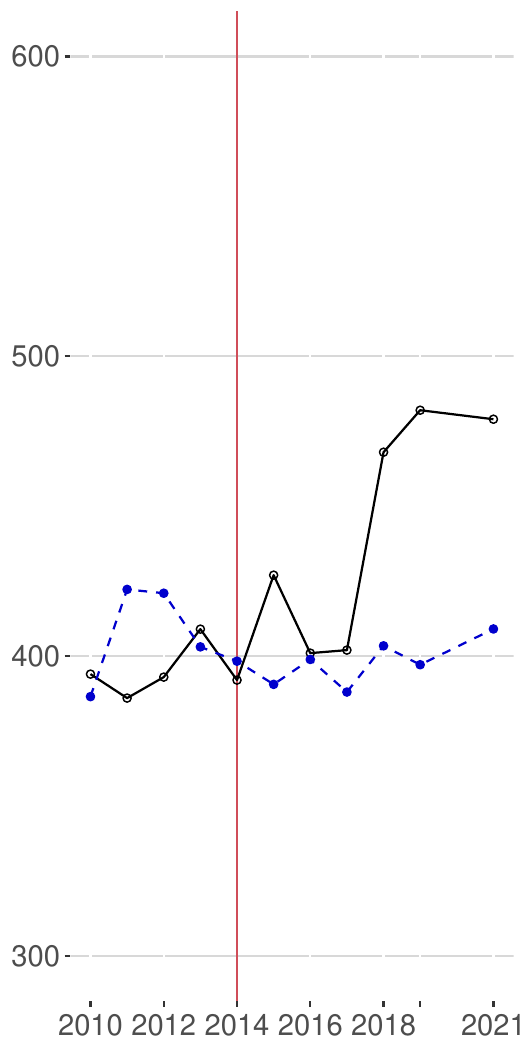}};
        \node[below=of img6, yshift=1cm] (img10) {\includegraphics[width=0.20\textwidth]{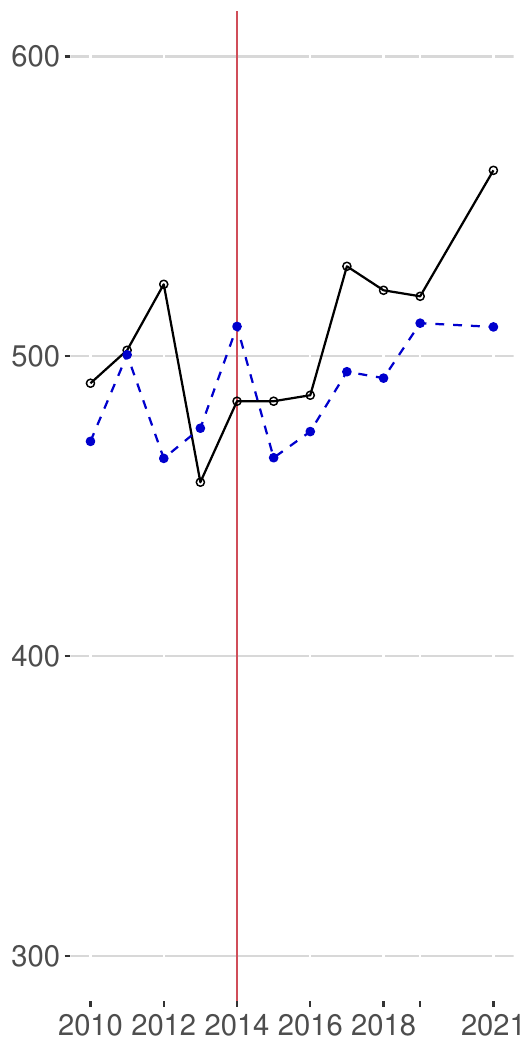}};
        \node[below=of img7, yshift=1cm] (img11) {\includegraphics[width=0.20\textwidth]{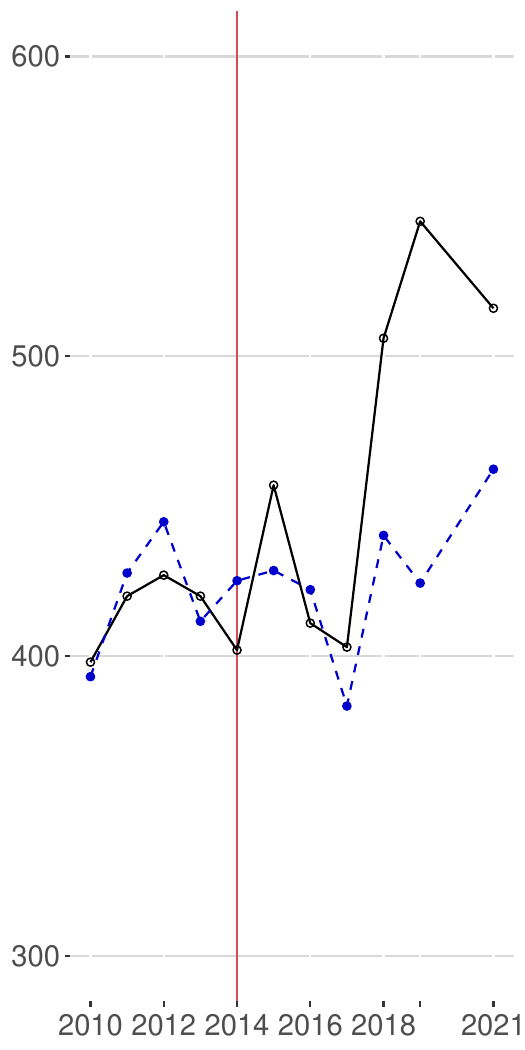}};
        \node[below=of img8, yshift=1cm] (img12) {\includegraphics[width=0.20\textwidth]{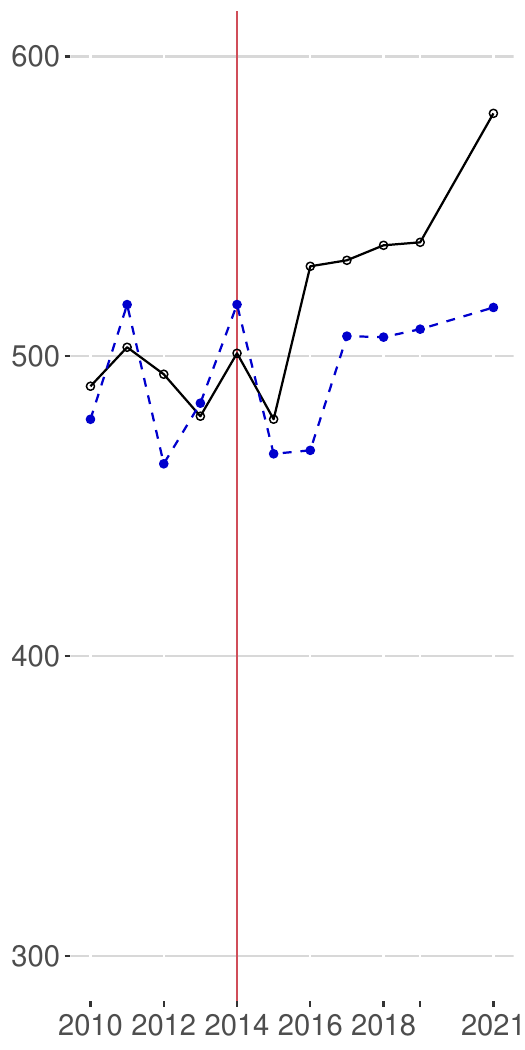}};

        \node[below=of img9, yshift=1cm] (img13) {\includegraphics[width=0.20\textwidth]{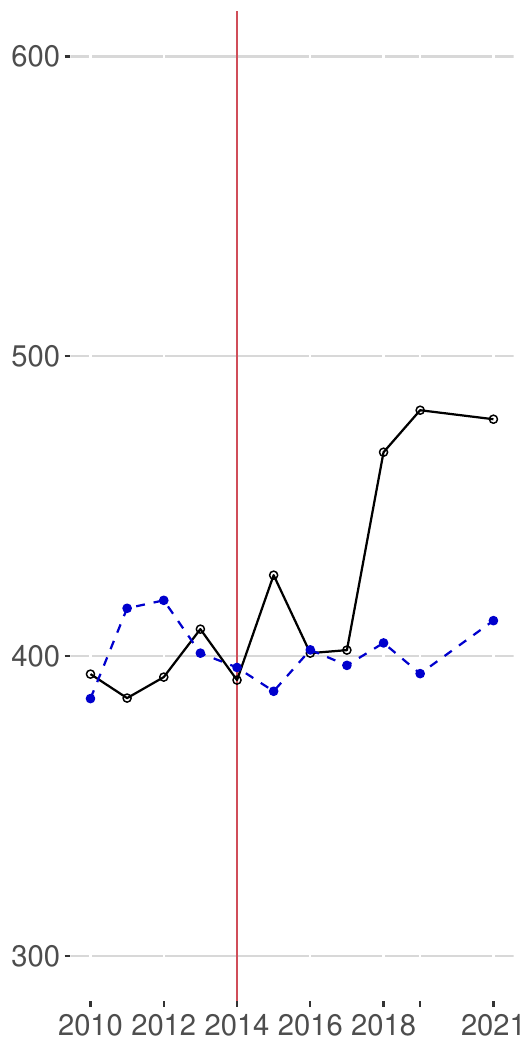}};
        \node[below=of img10, yshift=1cm] (img14) {\includegraphics[width=0.20\textwidth]{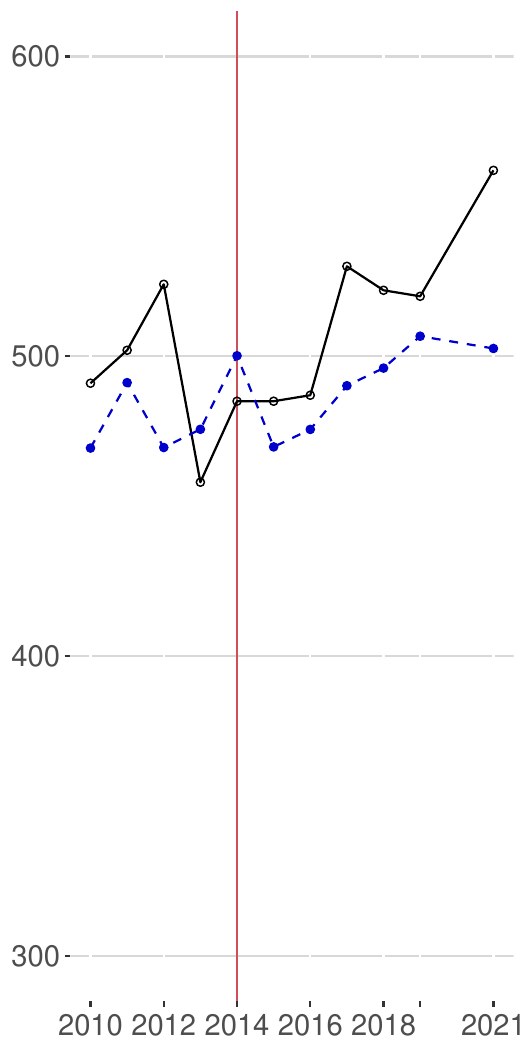}};
        \node[below=of img11, yshift=1cm] (img15) {\includegraphics[width=0.20\textwidth]{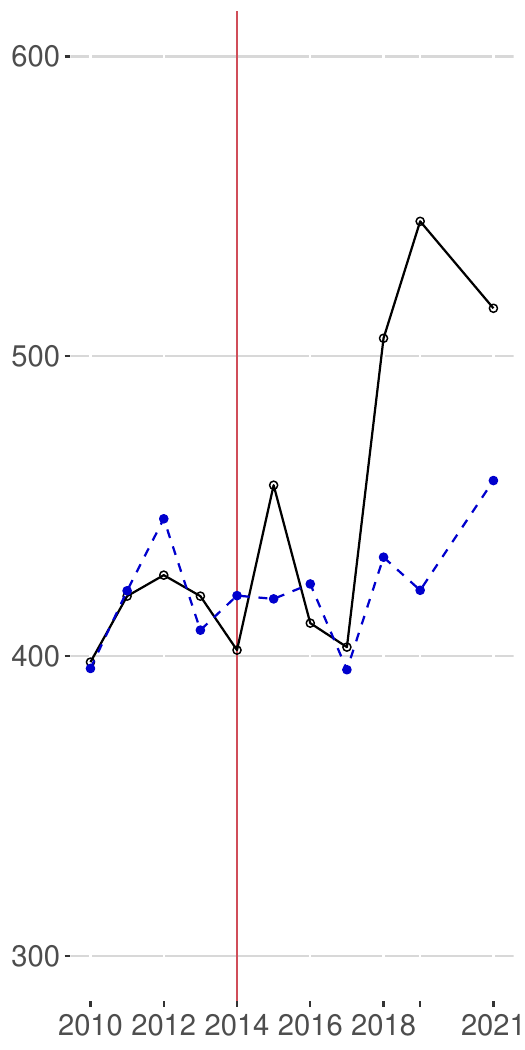}};
        \node[below=of img12, yshift=1cm] (img16) {\includegraphics[width=0.20\textwidth]{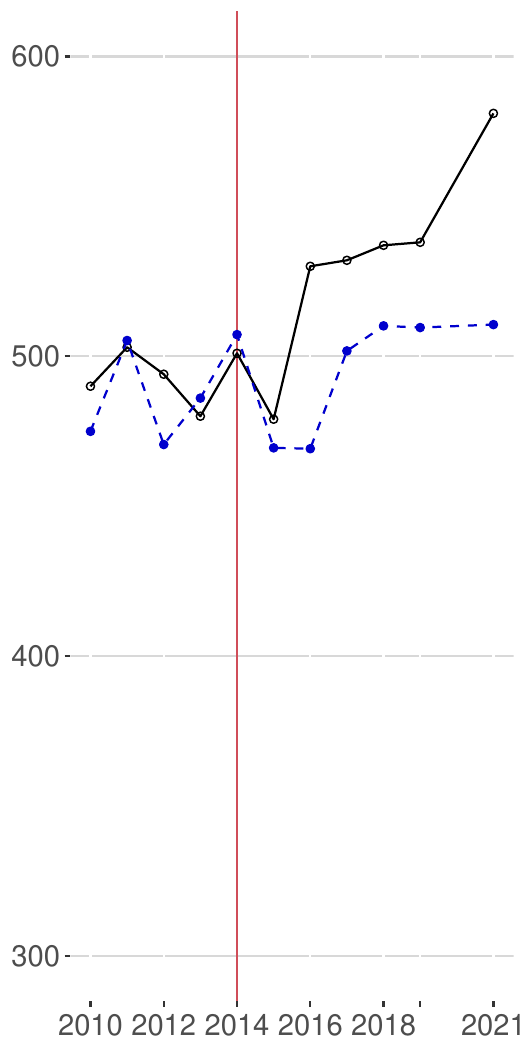}};

        \node[left=of img5, node distance=0cm, xshift=-0.5cm, rotate=90, anchor=center,yshift=-1cm,font=\color{black}] {\small Naplan score};
        \node[left=of img5, node distance=0cm, rotate=90, anchor=center,yshift=0cm,font=\color{black}] {Hillsborough};

        \node[left=of img9, node distance=0cm, xshift=-0.5cm, rotate=90, anchor=center,yshift=-1cm,font=\color{black}] {\small Naplan score};
        \node[left=of img9, node distance=0cm, rotate=90, anchor=center,yshift=0cm,font=\color{black}] {Carrington};

        \node[left=of img13, node distance=0cm, xshift=-0.5cm, rotate=90, anchor=center,yshift=-1cm,font=\color{black}] {\small Naplan score};
        \node[left=of img13, node distance=0cm, rotate=90, anchor=center,yshift=0cm,font=\color{black}] {Biraban};
        
    \end{tikzpicture}
    \caption{Leave-one-out estimates (continued)}
    \label{fig:loo1}
\end{figure}

\begin{figure}[htb]
    \centering
    \thispagestyle{empty}
    \begin{tikzpicture}
        \node (img1)  {\includegraphics[width=0.20\textwidth]{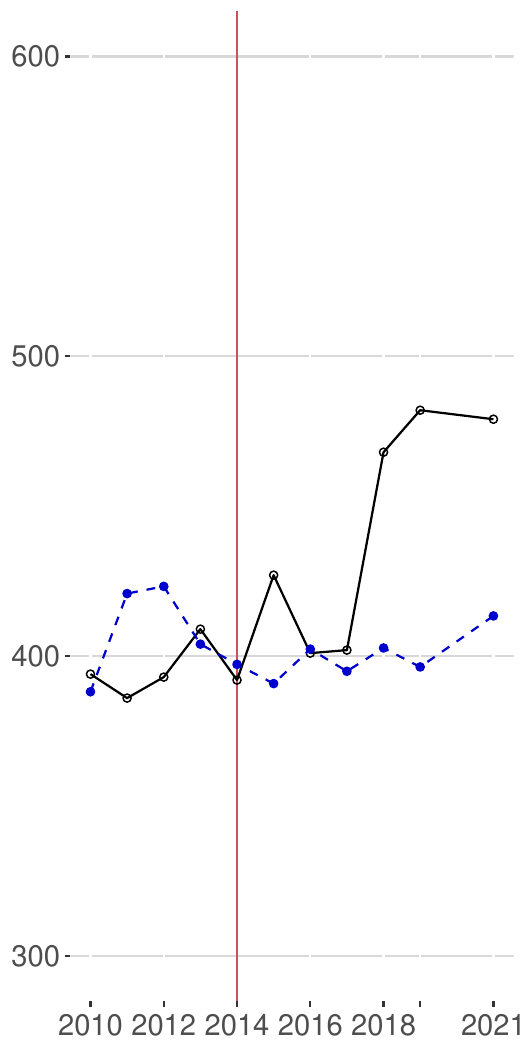}};
        \node[above=of img1, node distance=0cm, yshift=-1cm,font=\color{black}] {Numeracy Y-3};
        \node[left=of img1, node distance=0cm, xshift=-0.5cm, rotate=90, anchor=center,yshift=-1cm,font=\color{black}] {\small Naplan score};
        \node[left=of img1, node distance=0cm, rotate=90, anchor=center,yshift=0cm,font=\color{black}] {Stockton};

        \node[right=of img1, yshift=0cm, xshift=-1cm] (img2)  {\includegraphics[width=0.20\textwidth]{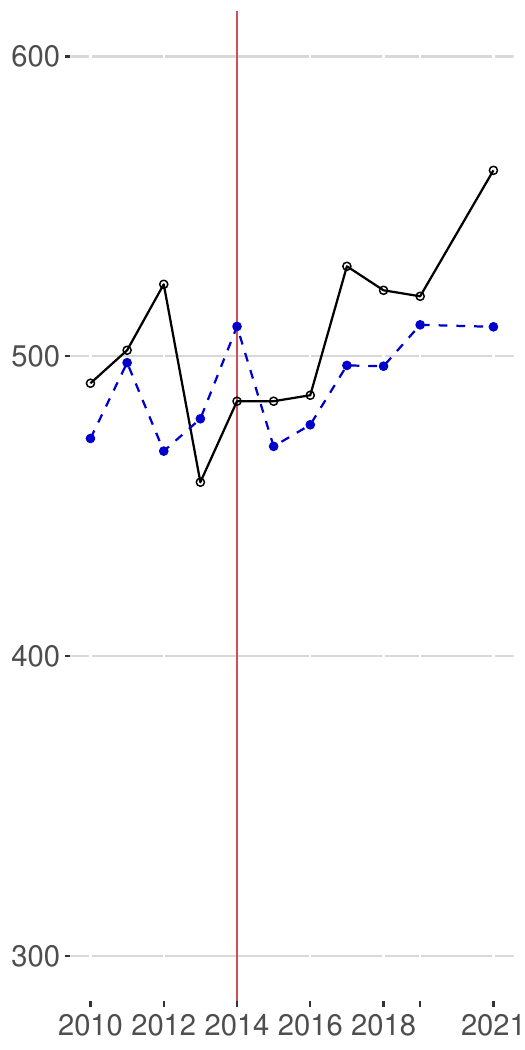}};
        \node[above=of img2, node distance=0cm, yshift=-1cm,font=\color{black}] {Numeracy Y-5};

        \node[right=of img2, yshift=0cm, xshift=-1cm] (img3)  {\includegraphics[width=0.20\textwidth]{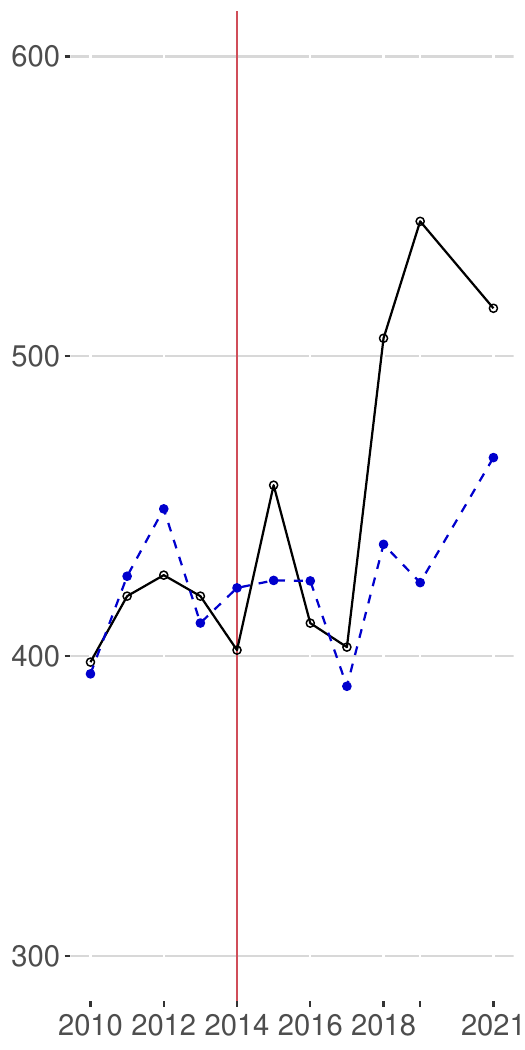}};
        \node[above=of img3, node distance=0cm, yshift=-1cm,font=\color{black}] {Reading Y-3};
        
        \node[right=of img3, yshift=0cm, xshift=-1cm] (img4)  {\includegraphics[width=0.20\textwidth]{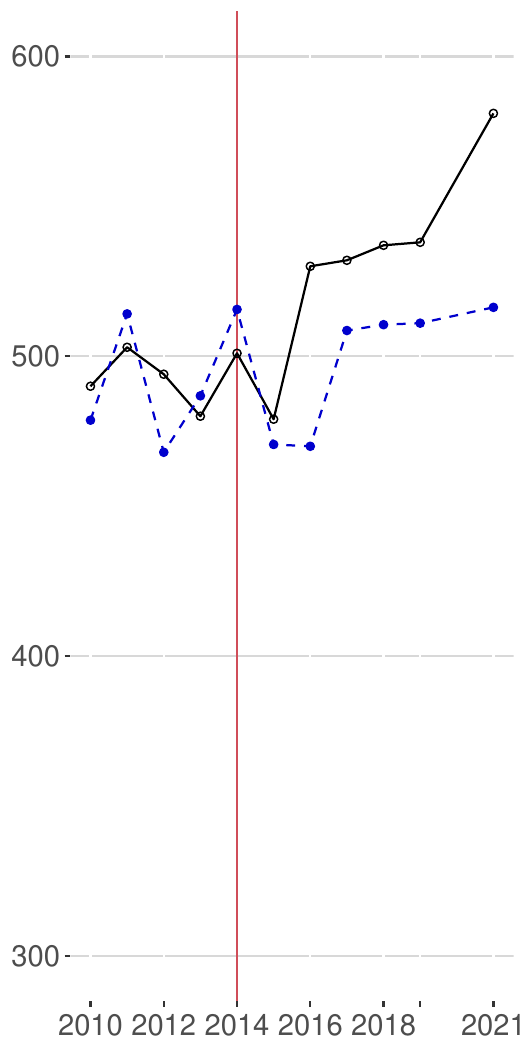}};
        \node[above=of img4, node distance=0cm, yshift=-1cm,font=\color{black}] {Reading Y-5};

        \node[below=of img1, yshift=1cm] (img5)  {\includegraphics[width=0.20\textwidth]{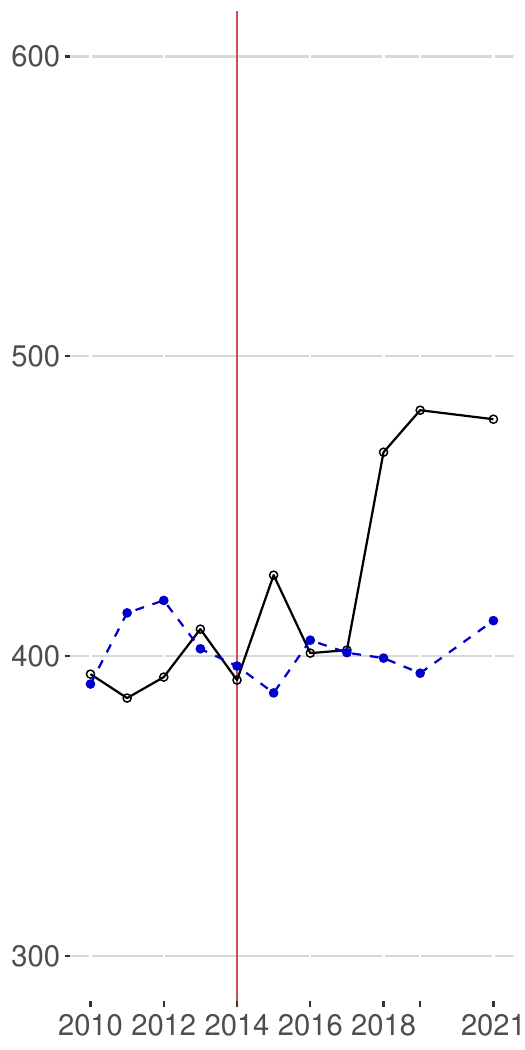}};
        \node[below=of img2, yshift=1cm] (img6)  {\includegraphics[width=0.20\textwidth]{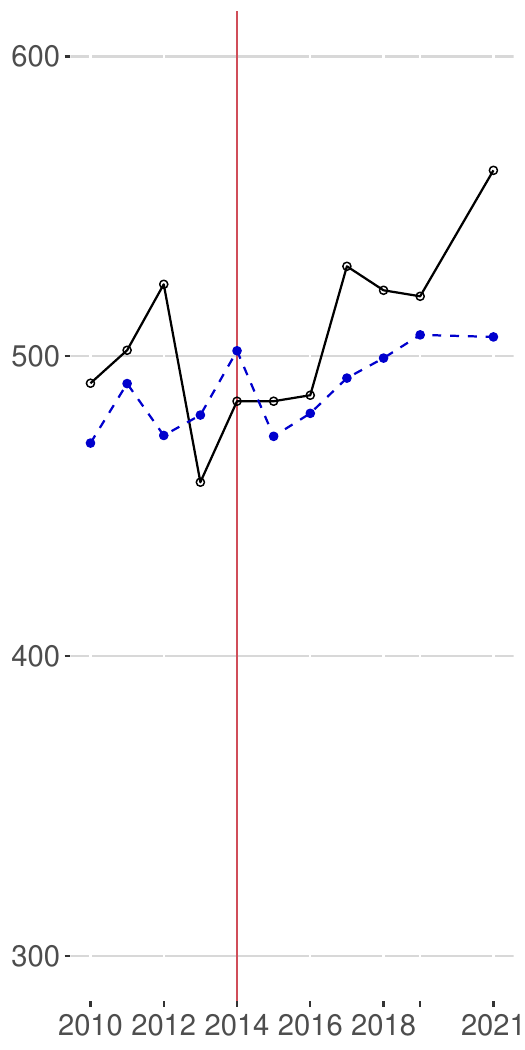}};
        \node[below=of img3, yshift=1cm] (img7)  {\includegraphics[width=0.20\textwidth]{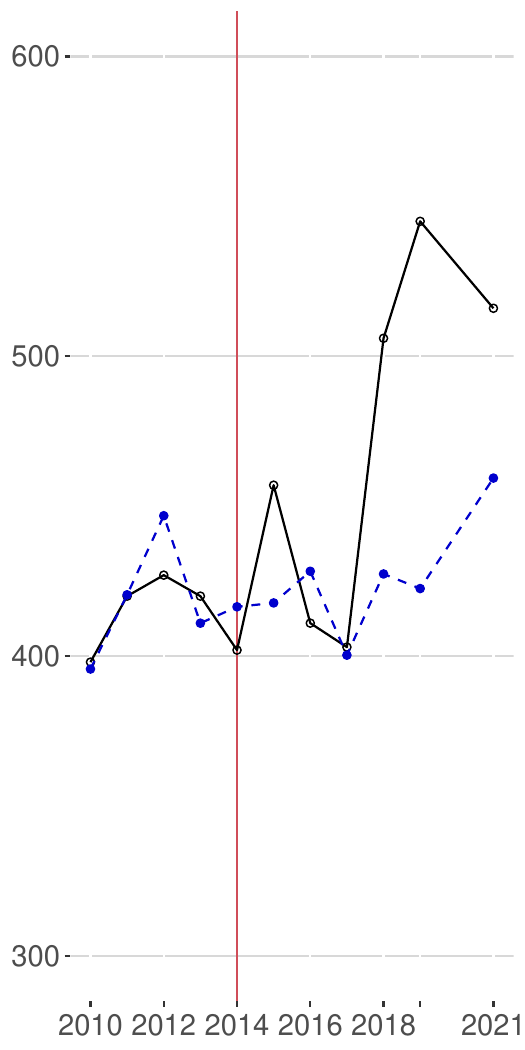}}; 
        \node[below=of img4, yshift=1cm] (img8)  {\includegraphics[width=0.20\textwidth]{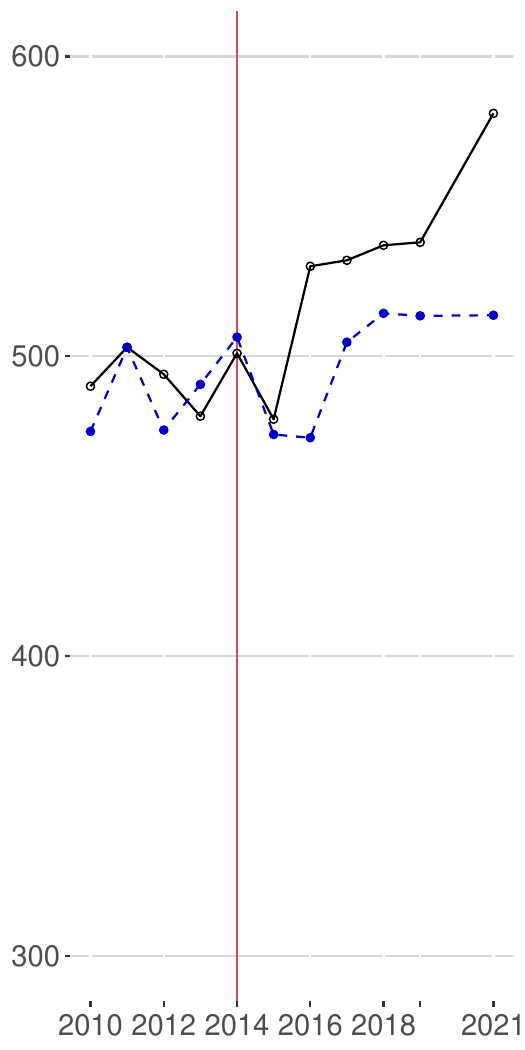}}; 

        \node[below=of img5, yshift=1cm] (img9)  {\includegraphics[width=0.20\textwidth]{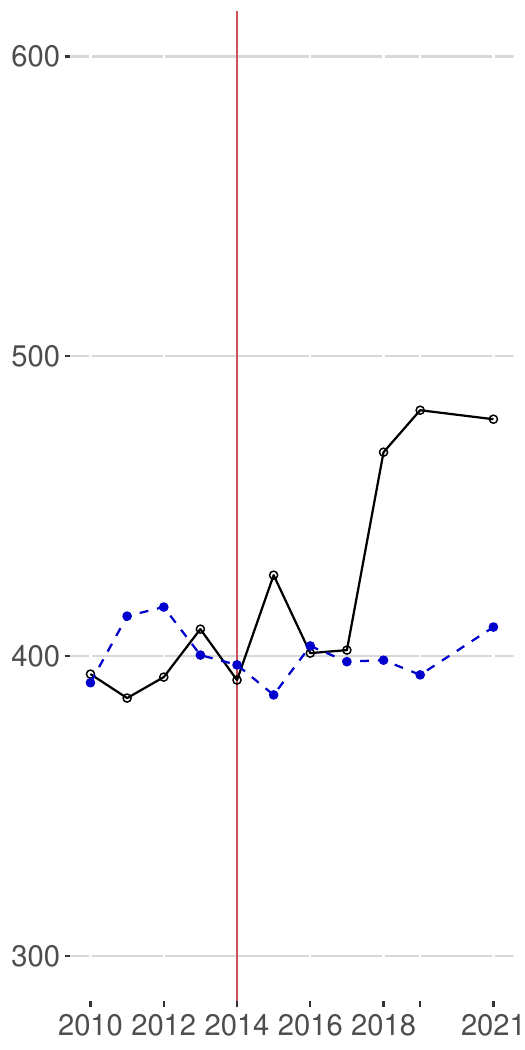}};
        \node[below=of img6, yshift=1cm] (img10) {\includegraphics[width=0.20\textwidth]{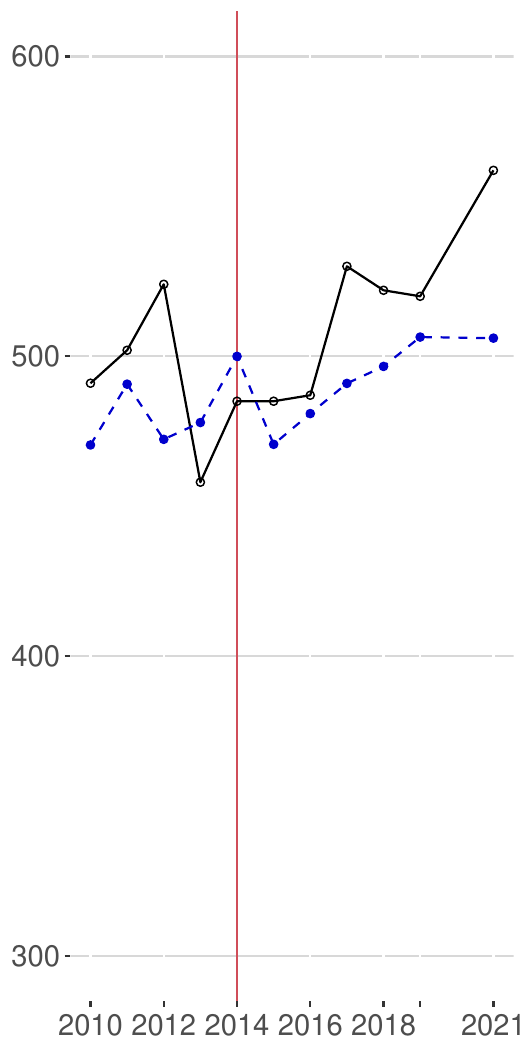}};
        \node[below=of img7, yshift=1cm] (img11) {\includegraphics[width=0.20\textwidth]{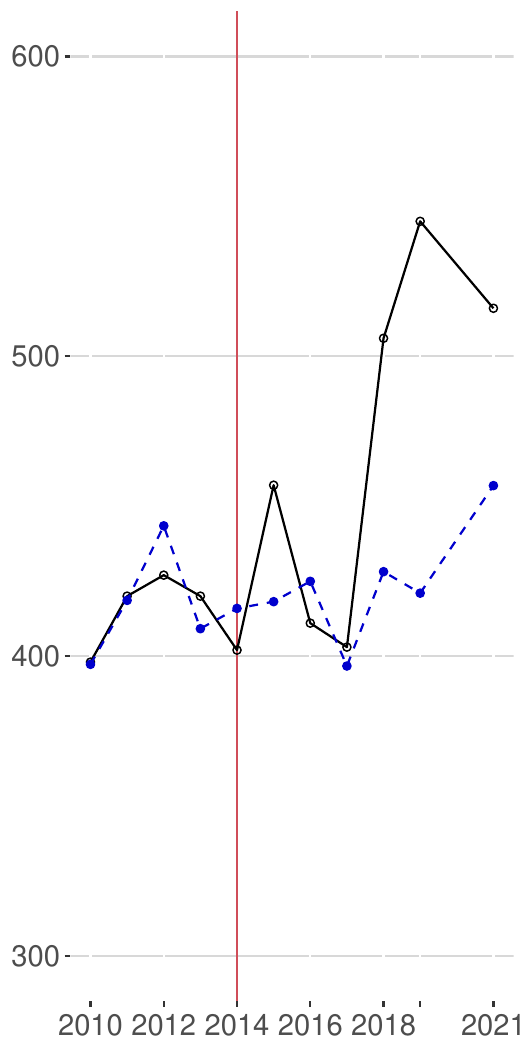}};
        \node[below=of img8, yshift=1cm] (img12) {\includegraphics[width=0.20\textwidth]{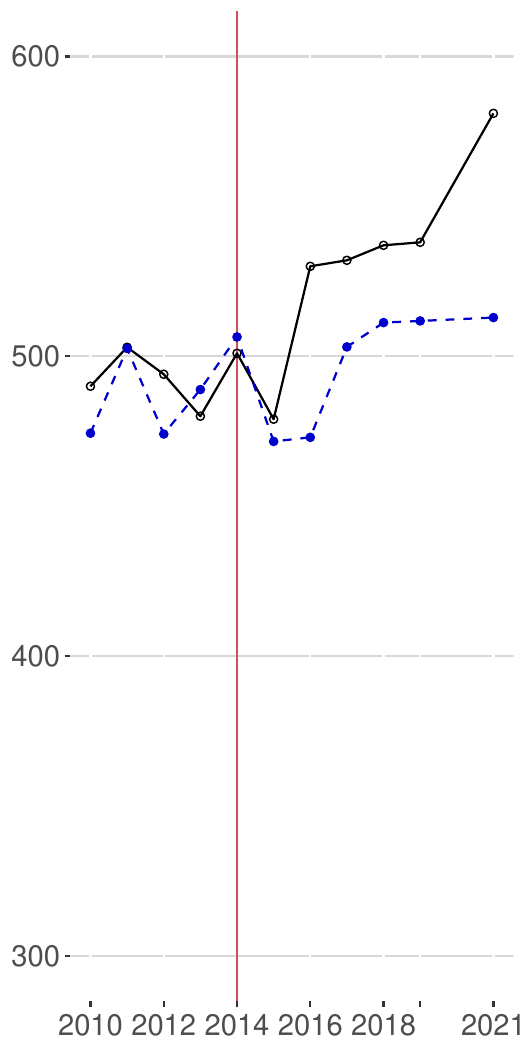}};

        \node[below=of img9, yshift=1cm] (img13) {\includegraphics[width=0.20\textwidth]{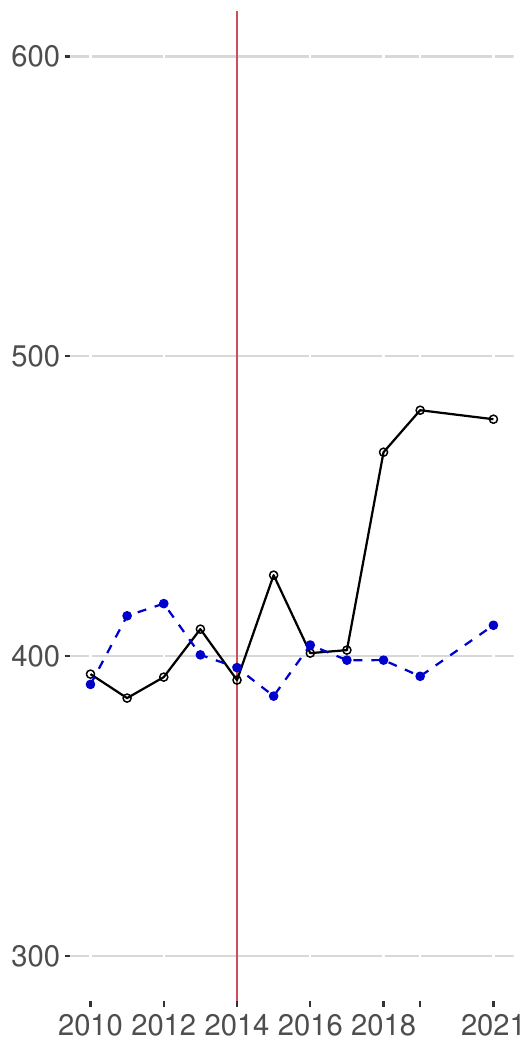}};
        \node[below=of img10, yshift=1cm] (img14) {\includegraphics[width=0.20\textwidth]{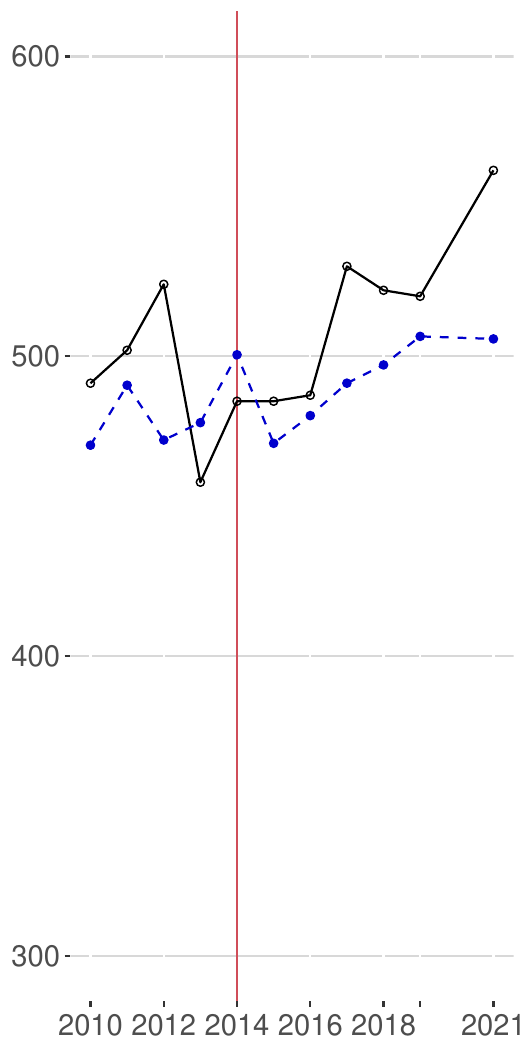}};
        \node[below=of img11, yshift=1cm] (img15) {\includegraphics[width=0.20\textwidth]{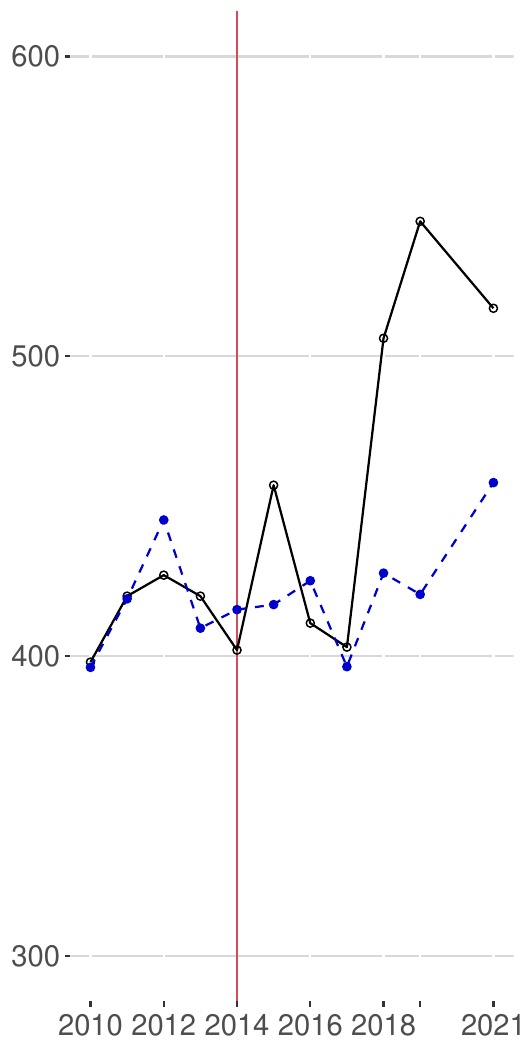}};
        \node[below=of img12, yshift=1cm] (img16) {\includegraphics[width=0.20\textwidth]{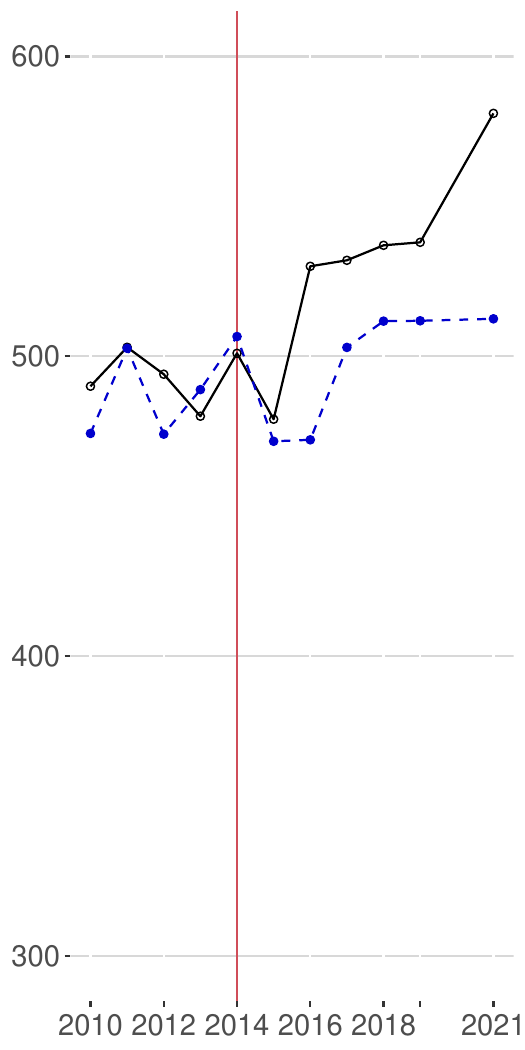}};

        \node[left=of img5, node distance=0cm, xshift=-0.5cm, rotate=90, anchor=center,yshift=-1cm,font=\color{black}] {\small Naplan score};
        \node[left=of img5, node distance=0cm, rotate=90, anchor=center,yshift=0cm,font=\color{black}] {Farmborough};

        \node[left=of img9, node distance=0cm, xshift=-0.5cm, rotate=90, anchor=center,yshift=-1cm,font=\color{black}] {\small Naplan score};
        \node[left=of img9, node distance=0cm, rotate=90, anchor=center,yshift=0cm,font=\color{black}] {Merewether};

        \node[left=of img13, node distance=0cm, xshift=-0.5cm, rotate=90, anchor=center,yshift=-1cm,font=\color{black}] {\small Naplan score};
        \node[left=of img13, node distance=0cm, rotate=90, anchor=center,yshift=0cm,font=\color{black}] {Coledale};
        
    \end{tikzpicture}
    \caption{Leave-one-out estimates}
    \label{fig:loo2}
\end{figure}

\newpage

\begin{table}[H]
\centering
\caption{\label{tab:LOO}Leave-One-Out estimates}
\centering
\begin{threeparttable}
\begin{tabular}[t]{llll}
\toprule
Year-3 Numeracy & Year-5 Numeracy & Year-3 Reading & Year-5 Reading\\
\midrule
74.28 & 31.88 & 85.87 & 39.95\\
(0.97) & (0.76) & (2.39) & (0.48)\\
{}[71.98, 76.58] & {}[30.09, 33.67] & {}[80.2, 91.53] & {}[38.82, 41.09]\\
\bottomrule
\end{tabular}
\begin{tablenotes}
\item Notes: 
  We run our main specification 8 times, leaving out in turn one of the top-8 
  donor schools for our main specification (See Table \ref{tab:weights}). 
  This table reports the resulting aggregated average treatment effect 
  estimates. We focus on the average treatment effect strictly three years 
  after the treatment date, as in the main text.
\end{tablenotes}
\end{threeparttable}
\end{table}

\begin{landscape}
\begin{table}
\centering
\caption{\label{tab:pctile}Performance percentiles of Charlestown South and donor schools}
\centering
\begin{threeparttable}
\begin{tabular}[t]{lrrrrrrrrrrrr}
\toprule
\multicolumn{1}{c}{ } & \multicolumn{3}{c}{Reading Y-3} & \multicolumn{3}{c}{Reading Y-5} & \multicolumn{3}{c}{Numeracy Y-3} & \multicolumn{3}{c}{Numeracy Y-5} \\
\cmidrule(l{3pt}r{3pt}){2-4} \cmidrule(l{3pt}r{3pt}){5-7} \cmidrule(l{3pt}r{3pt}){8-10} \cmidrule(l{3pt}r{3pt}){11-13}
School name & Pre & Post & Diff. & Pre & Post & Diff. & Pre & Post & Diff. & Pre & Post & Diff.\\
\midrule
Charlestown South Public School  \vspace{4pt}& 54 & 85 & \textbf{31} & 56 & 82 & \textbf{26} & 53 & 86 & \textbf{33} & 62 & 79 & \textbf{17}\\
Charlestown Public School& 51 & 68 & 17 & 52 & 59 & 7 & 71 & 57 & -14 & 46 & 67 & 21\\
Hillsborough Public School & 76 & 55 & -21 & 56 & 66 & 10 & 76 & 53 & -23 & 50 & 69 & 19\\
Carrington Public School & 20 & 40 & 20 & 21 & 36 & 15 & 27 & 58 & 31 & 14 & 45 & 31\\
Biraban Public School & 24 & 16 & -8 & 18 & 43 & 25 & 37 & 17 & -20 & 27 & 40 & 13\\
Stockton Public School & 42 & 49 & 7 & 45 & 52 & 7 & 47 & 52 & 5 & 43 & 50 & 7\\
Farmborough Road Public School & 26 & 16 & -10 & 14 & 17 & 3 & 13 & 22 & 9 & 7 & 8 & 1\\
Merewether Public School & 91 & 69 & -22 & 86 & 61 & -25 & 84 & 76 & -8 & 75 & 57 & -18\\
Coledale Public School & 91 & 92 & 1 & 81 & 82 & 1 & 77 & 81 & 4 & 76 & 72 & -4\\
\bottomrule
\end{tabular}
\begin{tablenotes}
\item \textit{Notes: } This table reports rankings of NAPLAN scores in 
  percentiles for Charlestown and its donor schools. 
  The rankings are calculated for each school, subject and year (3 and 5) using 
  mean NAPLAN scores before (2010-2013) and after (2014-2021) 
  the introduction of explicit instruction in Charlestown South. 
  Then, we take the difference between after and before 
  percentile, thus showing how many positions have been gained or lost 
  by each school in each subject after 2014.
\end{tablenotes}
\end{threeparttable}
\end{table}

\end{landscape}

\begin{figure}
        \centering
        \includegraphics[width=\textwidth]{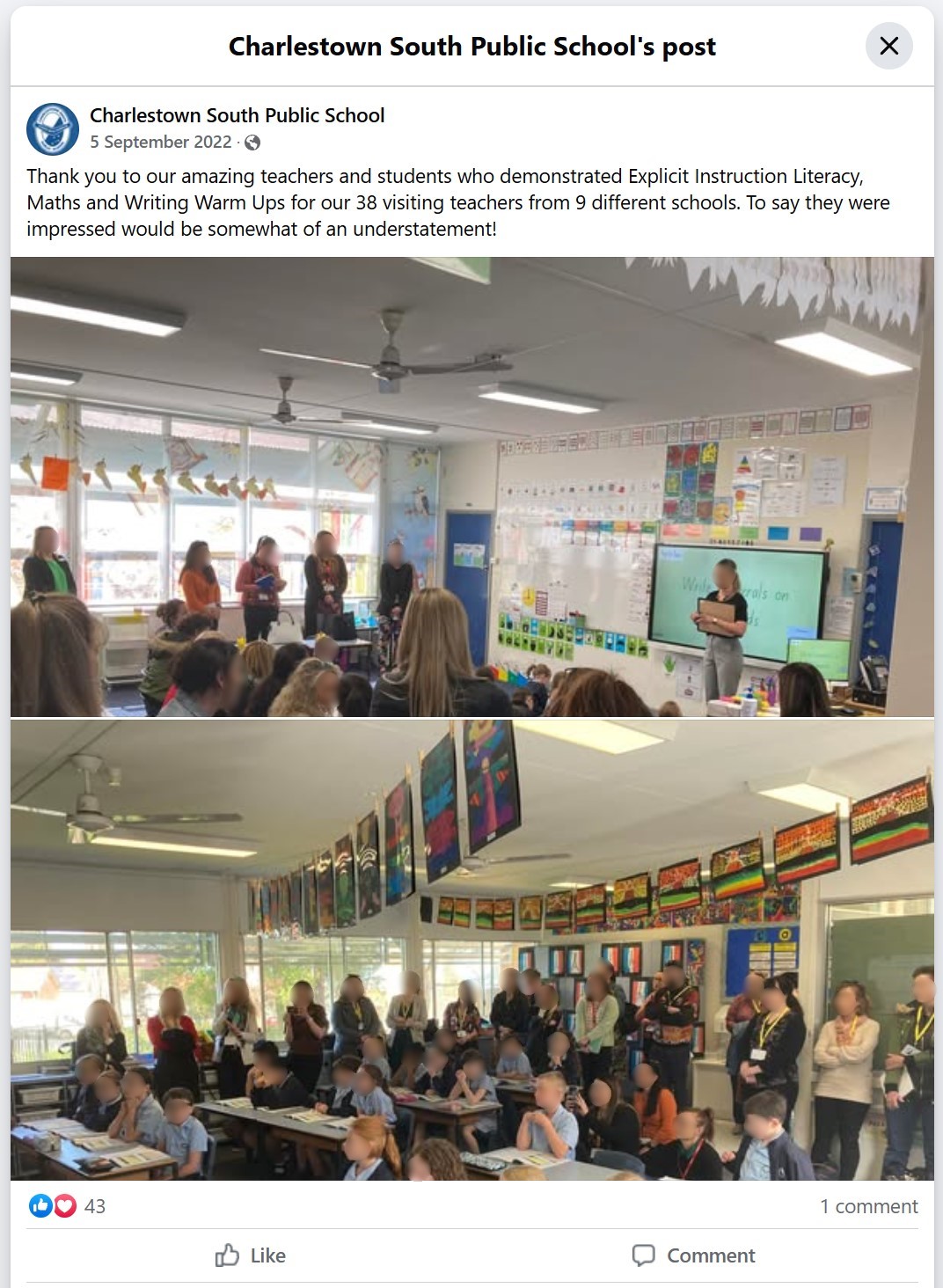}
    \caption{Facebook post}
    \label{fig:fb}
\end{figure}

\end{document}